\documentclass[10pt,twocolumn]{article}
\usepackage[a4paper,top=2.2cm,bottom=2.4cm,left=2.4cm,right=2.1cm,columnsep=6.5mm]{geometry}
\usepackage{iftex}
\ifXeTeX
  \usepackage{newtxtext,newtxmath}
  \usepackage{fontspec}
\else
  \usepackage[T1]{fontenc}
  \usepackage[helvratio=0.92]{newtxtext}
  \usepackage{newtxmath}
\fi
\usepackage{amsmath,graphicx,booktabs,textcomp,microtype,xcolor,enumitem}
\definecolor{accent}{HTML}{0B4F8A}
\definecolor{absbox}{gray}{0.93}
\usepackage[super,sort&compress,comma]{natbib}
\usepackage[colorlinks=true,allcolors=accent]{hyperref}
\usepackage{caption}
\DeclareCaptionLabelSeparator{pipe}{ $|$ }
\usepackage[nobottomtitles*]{titlesec}

\titleformat{\section}{\sffamily\bfseries\large\color{accent}}{\thesection}{0.55em}{}
\titleformat{\subsection}{\sffamily\bfseries\normalsize\color{accent}}{\thesubsection}{0.5em}{}
\titleformat{\subsubsection}{\sffamily\bfseries\small\color{accent}}{\thesubsubsection}{0.5em}{}
\titlespacing*{\section}{0pt}{11pt}{4pt}
\titlespacing*{\subsection}{0pt}{8pt}{2pt}
\titlespacing*{\subsubsection}{0pt}{6pt}{1pt}
\setlist[itemize]{leftmargin=1.2em,itemsep=1pt,topsep=2pt}
\usepackage{fancyhdr}
\fancypagestyle{firstpage}{\fancyhf{}\fancyfoot[L]{\footnotesize\sffamily Dash and Arvanitis}\fancyfoot[R]{\footnotesize\sffamily\thepage}}
\begin{document}
\twocolumn[{%
\begin{flushleft}
{\sffamily\footnotesize\color{accent}\textbf{ARTICLE}\enspace$\cdot$\enspace PREPRINT\par}
\vspace{6pt}
{\sffamily\bfseries\fontsize{20}{24}\selectfont Sub-wavelength registration of skull-compensating acoustic holograms via nonlinear parametric array feedback\par}
\vspace{9pt}
{\large Pradosh P. Dash$^{1}$ and Costas D. Arvanitis$^{1,2}$\par}
\vspace{4pt}
{\footnotesize $^{1}$George W. Woodruff School of Mechanical Engineering, Georgia Institute of Technology, Atlanta, GA 30332, USA.
$^{2}$Wallace H. Coulter Department of Biomedical Engineering, Georgia Institute of Technology and Emory University, Atlanta, GA
30332, USA.\par}
\vspace{6pt}
{\color{accent}\rule{\textwidth}{1.2pt}}
\end{flushleft}
\vspace{2pt}
\noindent\colorbox{absbox}{\parbox{\dimexpr\textwidth-2\fboxsep\relax}{\vspace{5pt}\textbf{Realizing the potential of optimized acoustic holograms for applications like transcranial ultrasound requires sub-wavelength alignment accuracy (<1.5~mm at 1~MHz), which is difficult to achieve with current non-MRI optical neuro-navigation methods. To address the critical challenge of alignment, we leverage the parametric array (PA) signal as an alignment surrogate, a highly penetrative, low-frequency difference wave generated by the nonlinear mixing of two primary high-frequency beams. Simulations show that a misaligned skull-compensating lens increases the primary acoustic energy within the skull bone (volumetric energy trapping), which strengthens the nonlinear source of the PA signal. We also tested whether shear-mode conversion in bone contributes: 2D elastic simulations showed only a small (3--5\%) increase in shear-type motion with misalignment, and the transmitted ex vivo bursts did not lengthen as a shear path would require. Pseudo-sound and microbubble artifacts were examined and found unlikely to explain the signal, consistent with the PA signal arising from the intrinsic nonlinearity of the bone. Minimizing this PA signal yields a noninvasive real-time feedback mechanism that achieves sub-wavelength alignment accuracy, outperforming standard neuronavigation. Our findings present a robust framework that supports the design of simple, economical, and high-performance transcranial ultrasound systems for high-precision neuro-interventions.}\vspace{5pt}}}
\par\vspace{5pt}
{\sffamily\footnotesize\color{accent}\textbf{Keywords:}\color{black}\ acoustic holography, transcranial focused ultrasound,
parametric array, nonlinear acoustics, image-free registration\par}
\vspace{12pt}
}]
\thispagestyle{firstpage}

\section{Introduction}\label{sec:1}

Our ability to reconfigure this potentially disruptive technology for biomedical applications, such as transcranial ultrasound, also depends on accurately registering the holographic lens to the patient's anatomy, orientation, and position relative to the transducer/lens plane.\cite{maimbourg2018,jimenezgambin2024} For instance, complex hologram designs and pressure field topologies, which require a higher operation frequency (\ensuremath{\geq} 0.7~MHz) and skull-compensating lens topologies, are very sensitive to lens-skull misalignment \cite{andres2022}. Hence, high-quality registration (i.e., sub-wavelength accuracy <1.5~mm at 1~MHz) is required to preserve fidelity and targeting accuracy. Unfortunately, current non-MRI-based registration methods (e.g., standard neuronavigation) typically achieve an accuracy of only \ensuremath{\sim} 2~mm, which can lead to targeting errors of a few millimeters (i.e., 1-2 wavelengths) and reduced performance \cite{choi2024,chen2021,wei2013,pouliopoulos2020}. Therefore, robust registration strategies that accurately align the lens with the patient's skull anatomy are critical for designing cost-effective, portable transcranial ultrasound (TUS) systems for high precision (i.e., subwavelength) neuro-interventions.

To address this critical challenge of registration, we hypothesize that nonlinear acoustic effects can be leveraged for noninvasive lens-skull registration. The hypothesis was tested through modeling and experiment. A parametric array (PA) signal, the low-frequency tone produced when two high-frequency beams mix nonlinearly \cite{westervelt1963,dash2021}, serves as the registration metric. Our investigations suggest that a misaligned (aberrated) lens enhances finite-amplitude wave propagation effects within a highly nonlinear skull, yielding a strong PA signal that can penetrate the skull with minimal loss. Thus, minimizing the PA signal leads to an effective acoustic feedback mechanism for noninvasively aligning the holographic lens with the skull.

In the following sections, we summarize the rationale and theory behind hologram registration using nonlinear PA feedback, followed by simulation and experimental results, as well as a sensitivity analysis. They help us explore the limitations of this method and draw key insights for clinical applications.

\section{Results}\label{sec:2}

\subsection{Skull-compensating lens misalignment augments nonlinear wave propagation and parametric array signal}\label{sec:2.1}

\begin{figure*}[tp]
\centering
\includegraphics[width=\textwidth,height=0.62\textheight,keepaspectratio]{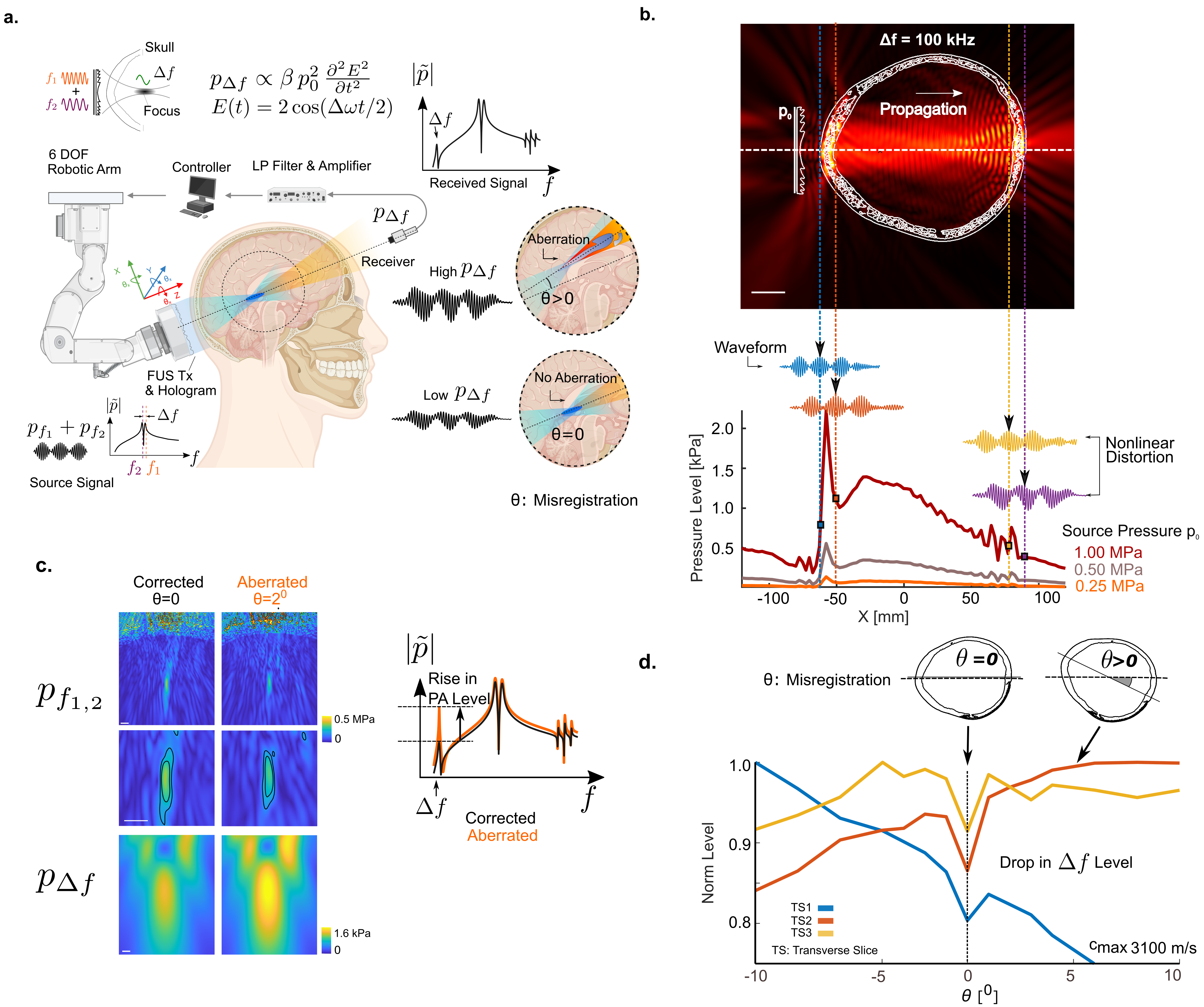}
\caption{\textbf{Conceptual framework of skull-compensating lens registration using nonlinear acoustic feedback.} \textbf{a}, Top: Nonlinear mixing of two high-frequency waves generated by a flat ultrasound (US) transducer attached to a phase plate, resulting in focused ultrasound. At higher intensities, this process produces a low difference frequency known as the parametric array (PA) effect. Bottom: Schematic illustration of the hologram-assisted FUS therapy device integrated with a six-degree-of-freedom (6 DOF) robotic arm. The system utilizes standard acoustic coupling (e.g., a gel or water bolus). A receiver is placed at a fixed, acoustically accessible location relative to the head. The system uses acoustic feedback from nonlinear parametric-array signals to achieve accurate registration. \textbf{b}, Top: 2D acoustic simulation demonstrating the generation of a 100~kHz parametric field within the skull cavity and its subsequent transmission. Bottom: Waveforms showing increasing nonlinear distortion as primary waves attenuate due to skull-induced losses, and absolute pressure traces along the axial direction for varying source pressures. \textbf{c}, Top: (Row 1), Simulation of the primary ultrasound field with 3~dB (in red) and 6~dB (in black) contour maps in the skull region showing local pressure maxima in the skull for skull rotation \ensuremath{\theta} = 0\textdegree{} and \ensuremath{\theta} = 2\textdegree{}, (Row 2) with zoomed view of the focal region as ROI with 3~dB and 6~dB contours (both in black); Bottom: The parametric field with (\ensuremath{\theta} = 2\textdegree{}) and without (\ensuremath{\theta} = 0\textdegree{}) skull rotation indicating aberration leads to higher PA signal (plot on the right). \textbf{d}, 2D simulation results illustrate a decrease in parametric pressure corresponding to zero skull rotation (\ensuremath{\theta} = 0\textdegree{}) for various transverse skull slices.}
\label{fig:1}
\end{figure*}

Given that high-frequency acoustic focusing is extraordinarily sensitive to registration errors, we addressed this long-standing challenge by turning our attention to largely unexplored non-linear wave propagation through the human skull to develop an active acoustic feedback mechanism. Briefly, the parametric array effect is a nonlinear wave-propagation phenomenon\cite{westervelt1963} in which two (primary) high-frequency sound beams of finite amplitude interact to produce (secondary) sum- and difference-frequency beams. The nonlinear mixing of two primary high-frequency beams generates a highly penetrative, low-frequency difference wave. The strength of this difference frequency |\ensuremath{p_{\Delta f}}|, here termed a parametric array (PA) signal, serves as a highly sensitive alignment surrogate. The PA signal, which has several unique properties, including high directionality and penetration through the skull, is proportional to the medium's nonlinearity parameter \ensuremath{\beta}, the square of the primary beams' amplitude \textit{p\textsubscript{f}}\textsubscript{1,2}, and their propagation length \cite{hamilton1997}. Considering the characteristics of the PA signal, we hypothesized that skull-compensating lens mis-registration (i.e., suboptimal aberration correction) can affect the amplitude of |\ensuremath{p_{\Delta f}}| that once detected and quantified (e.g., using a hydrophone) can provide a real-time feedback mechanism to noninvasively align the holographic lens to the patient's skull (Fig.~\ref{fig:1}a). This alignment procedure involves optimizing the position and orientation of the static lens relative to the skull prior to the TUS procedure to ensure accurate delivery of the intended acoustic field. Registration is performed before sonication by adjusting the lens pose to minimize the PA signal.

\begin{figure*}[tp]
\centering
\includegraphics[width=\textwidth,height=0.74\textheight,keepaspectratio]{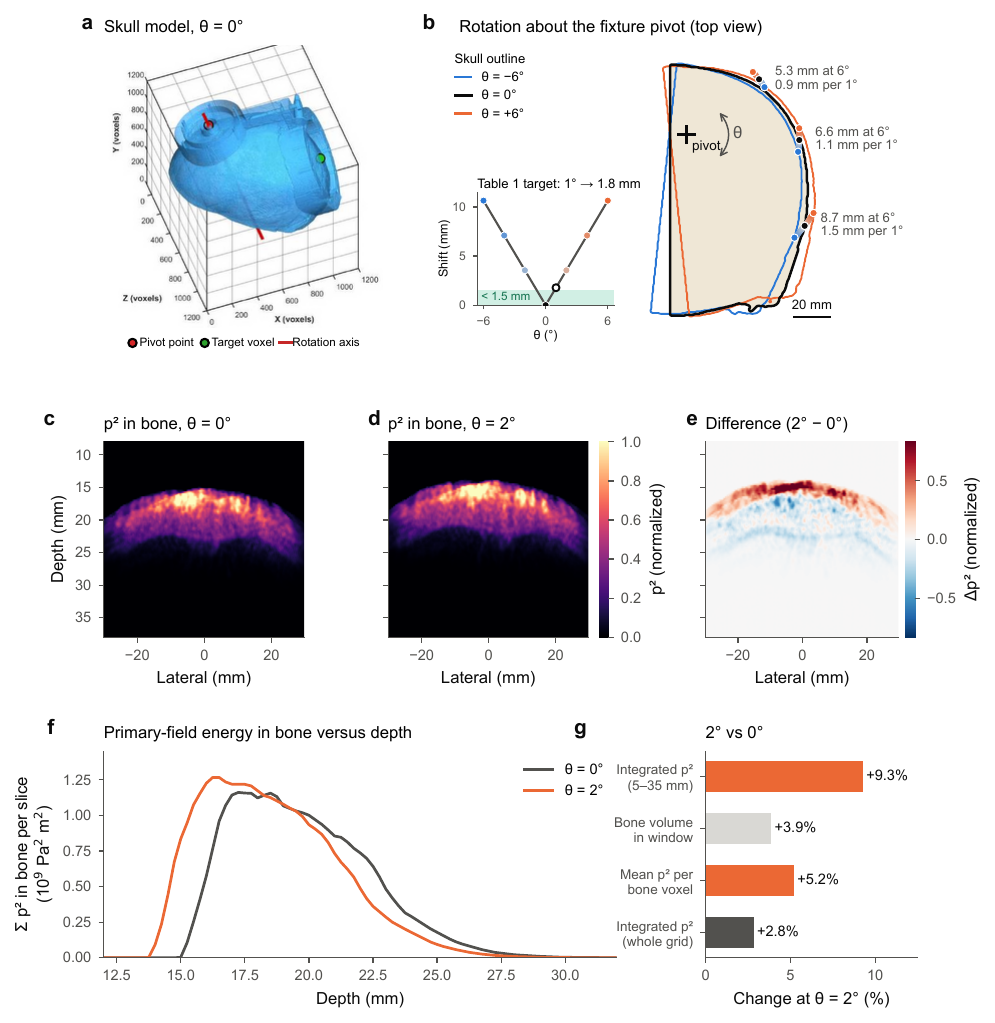}
\caption{\textbf{Kinematics of misregistration and primary-field energy in the skull.} \textbf{a}, Skull volume at \ensuremath{\theta} = 0\textdegree{} with the fixture pivot (red), its rotation axis and the target voxel (green) of Table~\ref{tab:1}. \textbf{b}, Top view of the models at \ensuremath{\theta} = \ensuremath{-}6\textdegree{}, 0\textdegree{} and +6\textdegree{}, viewed along the rotation axis. The pivot was fitted from the models: rotating the 0\textdegree{} model by \ensuremath{\pm}6\textdegree{} about it reproduces the \ensuremath{\pm}6\textdegree{} models (intersection over union 0.99). Coloured arcs trace three skull-surface points from \ensuremath{-}6\textdegree{} (blue) to +6\textdegree{} (orange); labels give their displacement between 0\textdegree{} and 6\textdegree{} and per 1\textdegree{} step. Inset, displacement of the Table~\ref{tab:1} target voxel (101.6~mm from the pivot) versus fixture rotation; the shaded band marks the 1.5~mm sub-wavelength target, which a single 1\textdegree{} step (open circle, 1.8~mm) already exceeds. \textbf{c},\textbf{d}, Squared peak primary pressure in bone voxels, summed across the out-of-plane direction, at \ensuremath{\theta} = 0\textdegree{} and 2\textdegree{} (3D simulations, B/A = 374), normalized to the 0\textdegree{} maximum. \textbf{e}, Difference between (d) and (c). \textbf{f}, Squared peak pressure in bone per depth slice. \textbf{g}, Relative changes from 0\textdegree{} to 2\textdegree{}: integrated squared pressure in bone at 5--35~mm (Table~\ref{tab:2}), bone volume in that window, mean per bone voxel, and integrated over the whole grid.}
\label{fig:2}
\end{figure*}

Understanding the necessity of this precise feedback loop requires characterizing the physical kinematics of transducer misalignment. During standard fixture rotation, misregistration occurs because the fixture rotates about a fixed vertical pivot axis P\textsubscript{pivot}, thereby spatially offsetting the transducer and inducing a coupled moment. A pure z-rotation, therefore, couples into lateral and axial translation. By tracking a target voxel on the skull surface (initially at T\textsubscript{initial}), we observe that the voxel traces a 3D arc through space. This kinematic setup translates a pure Z-axis rotation input into significant linear displacements in the lateral and depth directions (Fig.~\ref{fig:2}a,b). Assuming an isotropic voxel spacing of 150~\textmu m, geometric calculations (Table~\ref{tab:1}) confirm that a rotation of just \ensuremath{\pm}6\textdegree{} yields a geometric linear shift of approximately 10.6~mm. Even small angular errors produce large linear misregistration.

Nonlinear k-Wave simulations (Westervelt/PSTD) confirmed that the effect is detectable in a realistic skull geometry\cite{treeby2010}.

These models were run along with bone and tissue nonlinearity parameters from the literature\cite{renaud2008,pinton2011}. Using primary frequencies of 0.95~MHz and 1.05~MHz that resulted in 100~kHz difference frequency and pressures ranging from 0.25 to 1~MPa (safe exposure), we found that the parametric signal was immediately evident after the primary beams passed through the skull (Fig.~\ref{fig:1}b). Despite the formation of a weak standing wave (\ensuremath{\leq} 1.5~kPa), which is expected owing to the difference in frequency used \cite{baron2009}, the parametric signal outside the skull was also evident and well within the detection limits of many piezoelectric detectors (Fig.~\ref{fig:1} (b), bottom). While the observed temporal profile is atypical of nonlinear propagation (i.e., the peak positive pressure tends to be higher), this is due to the accumulation of nonlinearities over extended propagation distances, combined with the substantially higher attenuation of the primary and secondary MHz-range fields. In a clinical geometry the signal must cross two skull layers. The primary MHz tones are heavily attenuated on the return pass; the 100~kHz PA signal passes through with minimal loss (\ensuremath{\alpha} \ensuremath{\propto} \textit{f\textsuperscript{b}}). The skull itself acts as a low-pass filter, isolating the diagnostic tone.

\begin{table}[t]
\caption{\textbf{Displacement of the target voxel [1120, 620, 590] relative to the pivot [443, 644, 23] for fixture rotations of up to \ensuremath{\pm}6\textdegree{} (voxel size 150~\textmu m).} \ensuremath{\Delta}Z = 0 at all angles (rotation about the vertical axis).}
\label{tab:1}
\centering\small
\begin{tabular}{@{}lrrr@{}}
\toprule
Rotation (\textdegree{}) & \ensuremath{\Delta}X (mm) & \ensuremath{\Delta}Y (mm) & Displacement (mm) \\
\midrule
\ensuremath{-}6 & \ensuremath{-}0.18 & +10.63 & 10.64 \\
\ensuremath{-}4 & +0.00 & +7.09 & 7.09 \\
\ensuremath{-}2 & +0.06 & +3.55 & 3.55 \\
0 & 0.00 & 0.00 & 0.00 \\
+2 & \ensuremath{-}0.19 & \ensuremath{-}3.54 & 3.55 \\
+4 & \ensuremath{-}0.50 & \ensuremath{-}7.08 & 7.09 \\
+6 & \ensuremath{-}0.93 & \ensuremath{-}10.60 & 10.64 \\
\bottomrule
\end{tabular}
\end{table}

Replacing the intracranial medium with water (\ensuremath{\beta_{\mathrm{water}}} = 3.5) or brain (\ensuremath{\beta_{\mathrm{brain}}} = 4.45) changed the primary field negligibly but suppressed the PA signal when skull nonlinearity was absent. Bone (\ensuremath{\beta} \ensuremath{\approx} 188) overwhelms the volume integral in equation~(\ref{eq:2}); the skull region under the primary beam dominates PA generation.

\begin{table}[t]
\caption{\textbf{Primary-field energy in the skull.} Integrated squared primary pressure over bone voxels (c > 1700~m/s) at axial positions of 5 to 35~mm, which excludes the lens. Rotation also increases the bone volume in this window by 3.9\%; the mean squared pressure per bone voxel increases by 5.2\%.}
\label{tab:2}
\centering\small
\begin{tabular}{@{}lr@{}}
\toprule
Registration state & \begin{tabular}[b]{@{}r@{}}Integrated squared pressure\\in bone (Pa\textsuperscript{2}~m\textsuperscript{3})\end{tabular} \\
\midrule
Aligned (\ensuremath{\theta} = 0\textdegree{}) & 7.729 \ensuremath{\times} 10\textsuperscript{6} \\
Misaligned (\ensuremath{\theta} = 2\textdegree{}) & 8.446 \ensuremath{\times} 10\textsuperscript{6} \\
\midrule
Relative increase & +9.3\% \\
\bottomrule
\end{tabular}
\end{table}

Building on these observations, we tested the impact of lens misregistration on |\ensuremath{p_{\Delta f}}|. We found that, in the presence of misregistration, the PA signal increases substantially (Fig.~\ref{fig:1}c, right-hand plot). Further analysis revealed that the aberrated beams, in addition to distorting the pressure field (i.e., defocusing), also lead to a greater pressure buildup in the highly nonlinear skull.\cite{dash2025} The quantitative integration of the volumetric data revealed an accumulation of acoustic energy in the misaligned state. We established a baseline potential of 7.729\ensuremath{\times}10\textsuperscript{6} Pa\textsuperscript{2} m\textsuperscript{3} in the aligned state (\ensuremath{\theta} = 0\textdegree{}). Integration of the squared primary pressure within the skull bone volume revealed a +9.3\% increase for a 2\textdegree{} misalignment, increasing to 8.446\ensuremath{\times}10\textsuperscript{6} Pa\textsuperscript{2} m\textsuperscript{3} (Table~\ref{tab:2}, Fig.~\ref{fig:2}c--g). This increases the nonlinear source term (\ensuremath{S_{\mathrm{NL}}} \ensuremath{\propto} \ensuremath{\beta}|\ensuremath{p_{\mathrm{primary}}}|\textsuperscript{2}), consistent with the proposed feedback mechanism. Part of this increase reflects a 3.9\% larger bone volume within the integration window after rotation; the mean squared pressure per bone voxel increased by 5.2\% (Methods). At 2\textdegree{} the bone lies about 1~mm closer to the lens, so part of the increase reflects the skull moving into the stronger near field (Fig.~\ref{fig:2}e,f).

This effectively extends the nonlinear interaction region, which is critical for the development of finite-amplitude effects \cite{hamilton1997}. Finally, we assessed the influence of misregistration on the PA signal by rotating the skull (Fig.~\ref{fig:1}d). Interestingly, we found a steep increase in the PA signal for very small angles (i.e., small misregistration errors). Because the fixture rotates about an offset pivot, each rotation also translates the skull relative to the lens (up to 10.6~mm at \ensuremath{\pm}6\textdegree{}; Table~\ref{tab:1}), so these sweeps combine rotational and translational misregistration; pure translations were not simulated separately.

Crucially, these observations persisted for different skull slices (Fig.~\ref{fig:1}d), indicating that the PA signal drop is persistent and sensitive to the skull-compensating lens alignment. Together, these findings supported the notion that the low frequency acoustic signal generated by the nonlinear mixing of high frequency beams can be leveraged to attain accurate skull-compensating lens alignment.

\subsection{Sensitivity analysis reveals that parametric array signal is a robust and sensitive surrogate to skull-compensating lens alignment}\label{sec:2.2}

\begin{figure*}[tp]
\centering
\includegraphics[width=\textwidth,height=0.78\textheight,keepaspectratio]{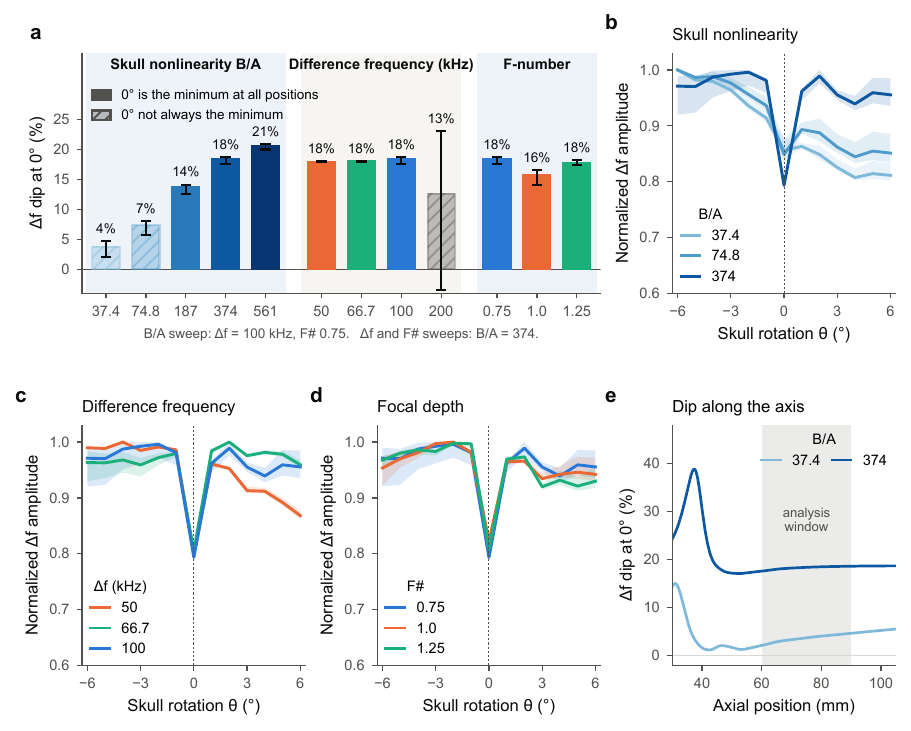}
\caption{\textbf{Sensitivity of the \ensuremath{\Delta}f dip at alignment (3D simulations, rotation about the z-axis).} \textbf{a}, Dip at 0\textdegree{} for every tested condition: skull B/A (\ensuremath{\Delta}f = 100~kHz, F\# 0.75), difference frequency and focal depth (B/A = 374). Bars: median over 30 axial positions between 60 and 90~mm from the grid origin (10--40~mm beyond the focus of the default lens); whiskers: minimum to maximum; solid bars: 0\textdegree{} is the minimum of the rotation curve at every position, hatched bars: it is not. \textbf{b}--\textbf{d}, Normalized \ensuremath{\Delta}f amplitude versus rotation for skull B/A, difference frequency and focal depth (F\# 0.75, 1.0 and 1.25 correspond to focal distances of 45, 60 and 75~mm with a 60~mm aperture); lines: mean over positions, bands: minimum to maximum. \textbf{e}, Dip along the beam axis for B/A = 37.4 and 374; shading marks the analysis window. Dip = 1 \ensuremath{-} A(0\textdegree{})/mean[A(\ensuremath{\pm}1\textdegree{})], with A normalized at each position to its maximum over \ensuremath{\theta} = \ensuremath{-}6\textdegree{} to 6\textdegree{}.}
\label{fig:3}
\end{figure*}

Motivated by these initial observations of volumetric energy trapping and augmented PA signals, we systematically evaluated the robustness of this metric using comprehensive three-dimensional (3D) nonlinear simulations. First, we investigated the impact of skull nonlinearity (\textit{B}/\textit{A})\textsubscript{skull} on parametric generation. As expected, the primary field (1.05~MHz) remained unaltered across different levels of skull nonlinearity; however, the parametric field (100~kHz) decreased markedly when skull nonlinearity was absent (Fig.~\ref{fig:5}c). To further clarify this observation, we varied the skull B/A (374, 74.8 and 37.4; Goldberg numbers 3.0, 0.62 and 0.32, respectively) and performed multiple registration iterations by rotating the 3D skull in the transverse plane. Evidently, the parametric pressure drop closely followed skull nonlinearity (Fig.~\ref{fig:3}a,b). Crucially, when skull misalignment is minimized (i.e., \ensuremath{\theta} = 0\textdegree{}), the drop in the PA signal becomes even more pronounced in the more realistic 3D simulations, as compared to 2D, for the same B/A parameters.

We also explored the effect of varying the difference frequency \ensuremath{\Delta}f (Fig.~\ref{fig:3}c) and observed that the PA signal drop appears to be insensitive to \ensuremath{\Delta}\textit{f} between 50 and 100~kHz (dip of about 18\% in each case); at 200~kHz the dip was smaller and less consistent (median 12.5\%, with 0\textdegree{} the minimum at 47\% of positions; Fig.~\ref{fig:3}a). Notably, the relationship between the PA signal amplitude and downshift ratio (\textit{f}/\ensuremath{\Delta}\textit{f}) follows established parametric array theory \cite{westervelt1963}, where larger downshift ratios yield smaller PA signals owing to lower nonlinear interaction efficiency. Conversely, smaller downshift ratios, while potentially producing stronger signals, require transducers with broader bandwidths and are hindered by higher frequency dependent attenuation through the propagation medium. This complex interplay of contributing factors determines the sensitivity of the PA signal changes to downshift-ratio variations. Next, we assessed the influence of focal depth by reducing the f-number while maintaining a constant aperture, producing a progressively weakly focused beam. The dip was similar for the three focal depths (16--18\%; Fig.~\ref{fig:3}d).

To assess clinical feasibility, we evaluated the dip at every point of a simulated region 40--105~mm deep and \ensuremath{\pm}30~mm wide beyond the skull (Fig.~S2). For a point receiver, 0\textdegree{} was the minimum of the rotation curve at 88\% of positions (median dip 22\%); averaging over a 20~mm receiver aperture raised this to 98\% and the 5th-percentile dip from 8\% to 14\%. Receiver positions elsewhere on the head were not simulated.

Across the tested difference frequencies and focal depths, the \ensuremath{\Delta}f amplitude at 0\textdegree{} was 16--18\% below its mean at \ensuremath{\pm}1\textdegree{} when the skull nonlinearity was high (B/A = 374); the dip fell to 4\% for B/A = 37.4 (Fig.~\ref{fig:3}a). For B/A = 374 the dip was stable along the axis beyond about 50~mm (Fig.~\ref{fig:3}e).

\subsection{Parametric array signal provides a real-time feedback mechanism to noninvasively align skull-compensating holographic lens to human skull}\label{sec:2.3}

\begin{figure*}[tp]
\centering
\includegraphics[width=\textwidth,height=0.66\textheight,keepaspectratio]{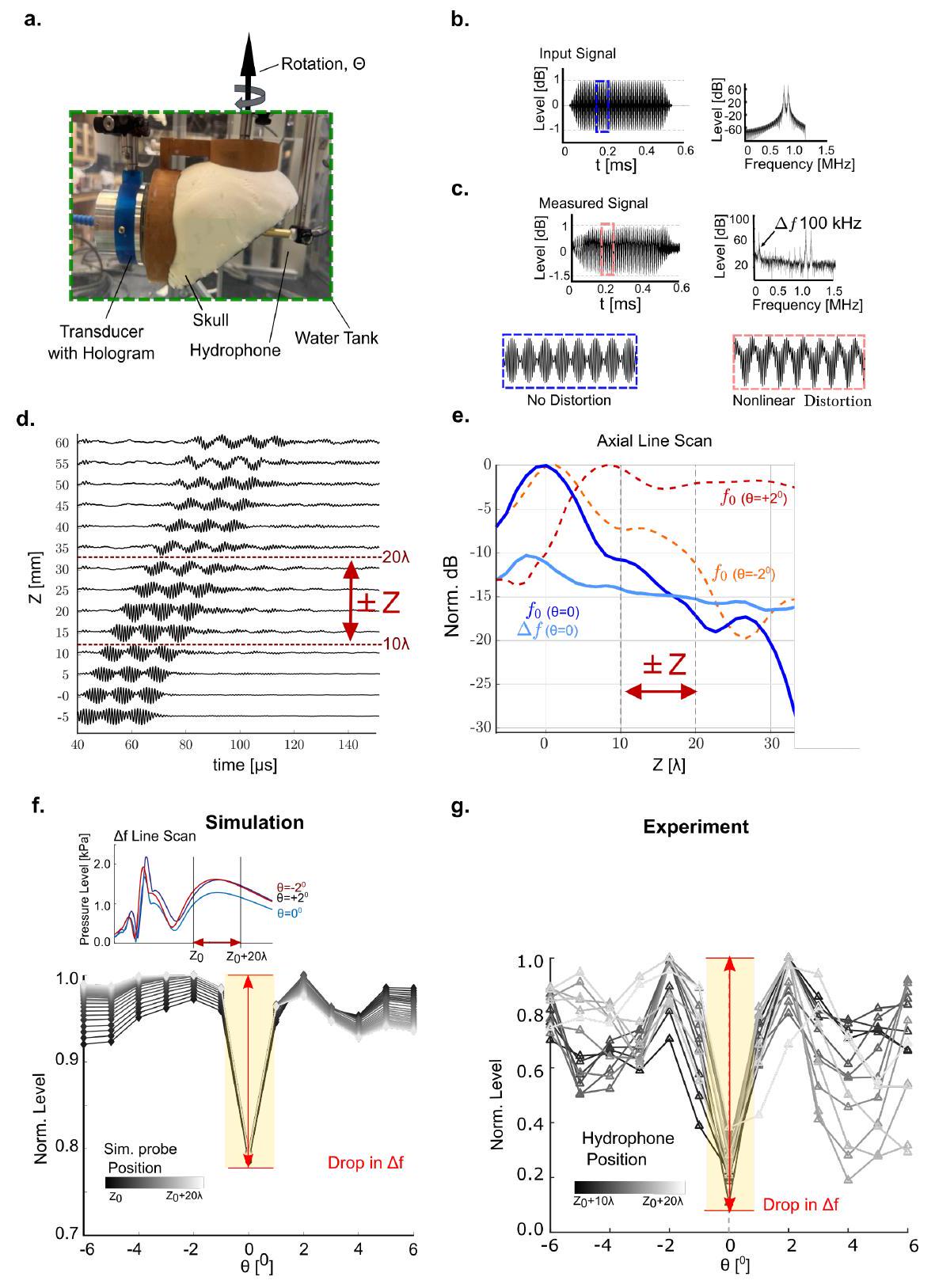}
\caption{\textbf{Low frequency acoustic feedback generated by nonlinear mixing of high frequency waves enables accurate skull-compensating lens alignment ex vivo.} \textbf{a}, Experimental setup with the skull mounted on a rotating fixture, with the skull cavity filled with degassed water. \textbf{b}, Sample bifrequency normalized input signal in both time and frequency domains; and Zoomed-in sections highlighting the absence of nonlinear distortion. \textbf{c}, Sample measured signals using hydrophone and after 600~kHz low pass filtering with 40~dB gain in both time and frequency domains; and Zoomed-in sections highlighting the presence of nonlinear distortion. \textbf{d}, Stacked waterfall plot indicating progressive nonlinear distortion of the measured signal (for \ensuremath{\theta} = 0\textdegree{}). \textbf{e}, Hydrophone line scans demonstrating Primary and parametric signals. \textbf{f}, 3D simulation mimicking the experimental setup, illustrating the variation of parametric pressure with skull rotation in the transverse plane. \textbf{g}, Experimental variation of parametric pressure, showing a decrease corresponding to zero registration error. In (f) and (g), normalization is performed across the different rotational angles (\ensuremath{\theta}) for measurements taken at specific axial positions (e.g., \ensuremath{Z_{0}}+20\ensuremath{\lambda}).}
\label{fig:4}
\end{figure*}

To experimentally validate the theoretical sensitivity of this nonlinear acoustic feedback mechanism, we designed a holographic lens using the HASA-ADAM framework\cite{dash2025a,arvanitis2025,dash2023} and conducted experiments with a 1~MHz transducer with an active aperture D = 60~mm, coupled with a single focusing lens (F\# 0.75). The transducer was excited with a bi-frequency input signal (containing 0.95~MHz and 1.05~MHz) at 0.2~MPa to produce a 100~kHz difference frequency. The excitation consisted of short tone bursts (30 cycles) to avoid standing wave artifacts. This signal was then recorded with a hydrophone and compared with three-dimensional simulations using the same geometry and skull segment (Fig.~\ref{fig:4}). To perform axial scans and characterize the PA signal at different distances from the skull using a hydrophone, we removed part of the skull (Fig.~\ref{fig:4}a). The skull cavity was immersed in degassed water to provide a standardized acoustic environment. This choice allowed us to isolate the effect of skull-induced aberration and nonlinearity on the PA signal, as water has low nonlinearity, and helped minimize the risk of pseudo sound artifacts due to hydrophone nonlinearity \cite{song2021}.

Axial line scans confirmed that a 0.2~MPa (peak-to-peak) primary field (peak pressure at the focus) produced increasing nonlinear distortion along the axis and beyond the focal position. This was evident in the waterfall plot showing progressive self-demodulation (Fig.~\ref{fig:4}d).

To minimize pseudo-sound effects that can appear in the measurements when the hydrophone is subject to strong primary pressure fields, the measurement window was extended several wavelengths away from the focus (Fig.~\ref{fig:4}d). We established the presence of a measurable PA signal for clinically relevant primary pressures (M.I. = 0.1), we rotated the skull at 1\textdegree{} increments around the z-axis and measured its amplitude, as in the simulations above (Figs.~\ref{fig:1} and~\ref{fig:3}). The line scans for the primary frequency revealed broadening of the axial beamwidth for both positive (\ensuremath{\theta} > 0\textdegree{}) and negative (\ensuremath{\theta} < 0\textdegree{}) registration errors (Fig.~\ref{fig:4}e). These measurements, aggregated across z-axis positions, not only closely aligned with the simulation predictions, but also confirmed the pronounced drop in parametric pressure when the skull rotation was zero (Fig.~\ref{fig:4}f--g). The normalization in Fig.~\ref{fig:4}g is performed across the different rotational angles (\ensuremath{\theta}) for measurements taken at specific axial positions (e.g., \ensuremath{Z_{0}}+20\ensuremath{\lambda}), not by spatial averaging. This procedure is accessible in practice; the operator monitors the signal at a fixed location while adjusting the lens orientation to find the minimum PA signal.

In aggregate, these findings (Figs.~\ref{fig:1}--\ref{fig:4}) support our hypothesis that the PA signal is sensitive to skull-induced aberration and demonstrate its potential to provide a real-time feedback mechanism to align the skull-compensating holographic lens to the patient's skull.

\subsection{Mechanisms Governing the Angular Sensitivity of Acoustic Feedback}\label{sec:2.4}

\begin{figure*}[tp]
\centering
\includegraphics[width=\textwidth,height=0.78\textheight,keepaspectratio]{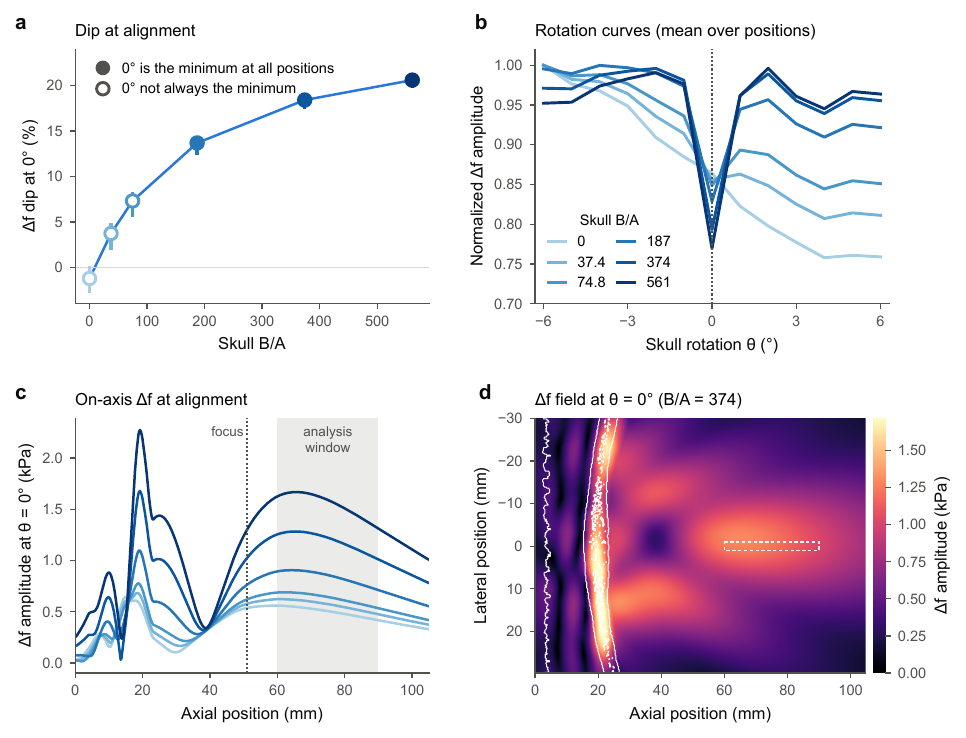}
\caption{\textbf{Effect of skull nonlinearity on the \ensuremath{\Delta}f dip (3D simulations, \ensuremath{\Delta}f = 100~kHz, F\# 0.75; same runs, window and metric as Fig.~\ref{fig:3}).} \textbf{a}, Dip at 0\textdegree{} versus skull B/A (median over 30 axial positions between 60 and 90~mm; bars: minimum to maximum; filled points: 0\textdegree{} is the minimum at every position). \textbf{b}, Normalized \ensuremath{\Delta}f amplitude versus rotation (mean over the same positions). \textbf{c}, On-axis \ensuremath{\Delta}f amplitude at \ensuremath{\theta} = 0\textdegree{}; shading marks the analysis window and the dashed line the focus. \textbf{d}, \ensuremath{\Delta}f amplitude at \ensuremath{\theta} = 0\textdegree{} for B/A = 374; white lines outline the lens and skull (sound speed above 1700~m/s) and the dashed box marks the analysis window.}
\label{fig:5}
\end{figure*}

In the preceding sections, the central result that a consistent PA drop at \ensuremath{\theta} = 0\textdegree{} in both simulations and experiment has been established. What follows investigates a secondary issue: experiments showed a steeper angular roll-off than the fluid model predicted. Three candidate explanations are mechanical fixture slop, an underestimated skull \textit{B}/\textit{A}, and unmodelled shear-wave generation. To evaluate the impact of the skull's nonlinearity on the registration metric, we quantified the normalized drop in the difference frequency (\ensuremath{\Delta}\textit{f}) across a range of \textit{B}/\textit{A} values (Fig.~\ref{fig:5}). The analysis reveals a direct, positive correlation: as the assumed nonlinearity of the skull increases, the magnitude of the signal drop at optimal alignment (\ensuremath{\theta} = 0\textdegree{}) becomes substantially larger. For a linear skull (B/A = 0), the \ensuremath{\Delta}f amplitude at 0\textdegree{} was not lower than at \ensuremath{\pm}1\textdegree{} (dip \ensuremath{-}1.2\%) and 0\textdegree{} was not the minimum of the rotation curve. The dip grew with B/A (3.7\%, 7.3\% and 13.7\% for B/A = 37.4, 74.8 and 187), and 0\textdegree{} became the minimum at all positions from B/A = 187 (Fig.~\ref{fig:5}). Increasing \textit{B}/\textit{A} to upper physiological estimates (B/A = 374) yields a dip of 18.4\%. This parametric relationship provides strong evidence that the steep angular drop-off observed in our ex vivo experiments could be largely attributed to a higher true \textit{B}/\textit{A} value of the physical skull segment compared to the conservative estimates used in our initial models. Consequently, underestimating the skull's nonlinearity stands as a highly probable, mechanically simple explanation for the discrepancy between simulation and experiment. Nevertheless, to examine all physical possibilities, we also explored mode conversion at the fluid-bone interface.

\textbf{Geometric incidence.} As a coarse indicator of mode conversion, we computed the fraction of skull-surface points inside the 60~mm beam cylinder whose normal makes more than 30\textdegree{} with the beam axis (Methods~\ref{sec:5.2}). This fraction was lowest at \ensuremath{\theta} = 0\textdegree{} (57\%) and rose in steps to 65--66\% beyond \ensuremath{\pm}4\textdegree{}. Because the metric assumes plane-wave incidence and includes internal trabecular surfaces, it indicates trends only; with the converging beam of an F\# 0.75 lens, the marginal rays already reach about 34\textdegree{} at alignment. Holograms that project complex patterns use steeper phase gradients and more oblique incidence, which would increase mode conversion further.

\begin{figure*}[tp]
\centering
\includegraphics[width=\textwidth,height=0.8\textheight,keepaspectratio]{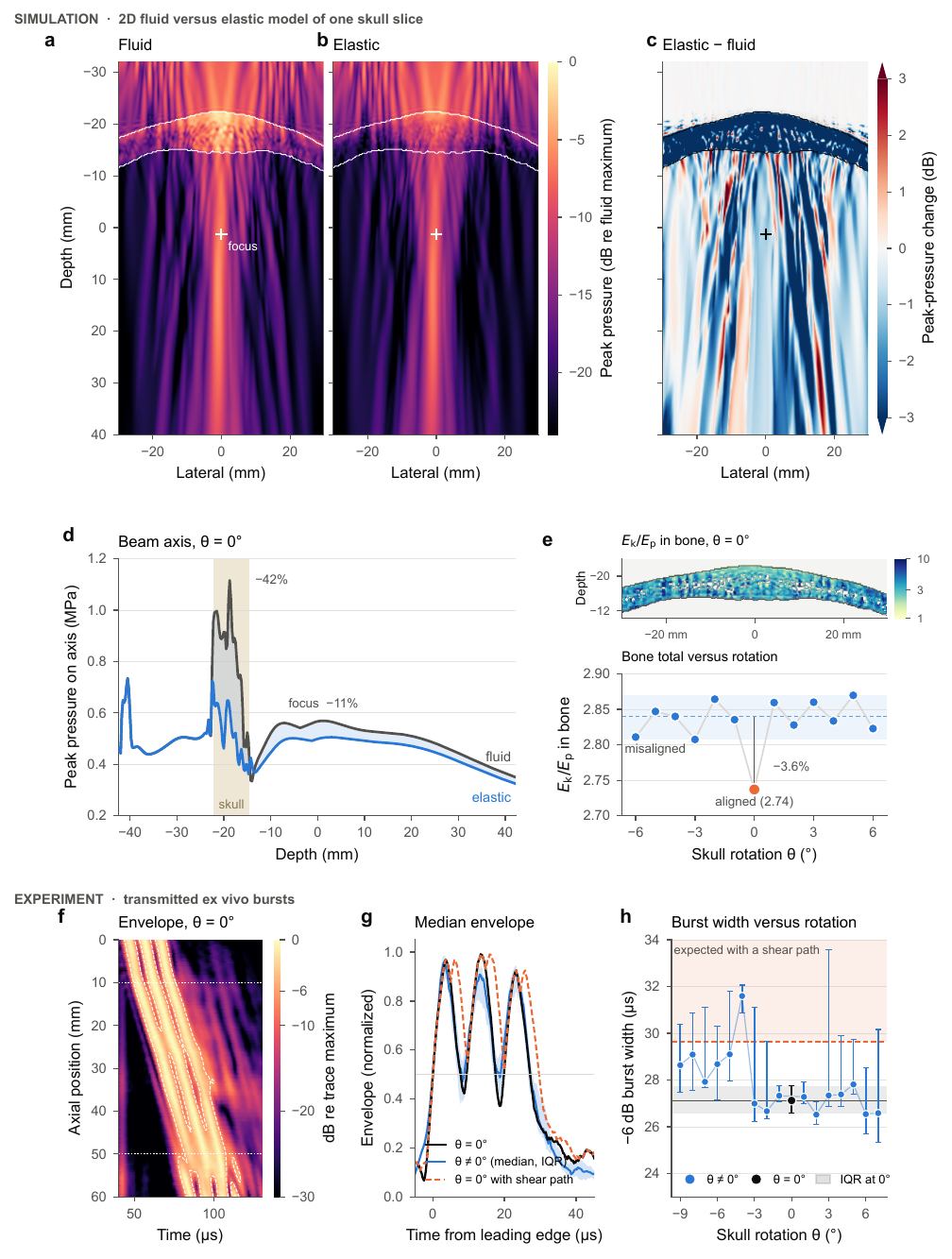}
\caption{\textbf{Testing the shear hypothesis.} \textbf{a}--\textbf{e}, 2D fluid and elastic simulations of one skull slice (CT-derived compressional sound speed and density; shear speed 1400~m/s in bone). \textbf{a},\textbf{b}, Peak pressure at \ensuremath{\theta} = 0\textdegree{} in dB relative to the fluid maximum; white lines, outer and inner skull surfaces; cross, focus. \textbf{c}, Change of the peak pressure when shear is included, elastic minus fluid in dB (clipped at \ensuremath{\pm}3~dB). \textbf{d}, Peak pressure on the beam axis at \ensuremath{\theta} = 0\textdegree{}; beige, skull; blue shading, pressure lost when shear is included (skull 1.12 to 0.65~MPa, \ensuremath{-}42\%; focus 0.57 to 0.51~MPa, \ensuremath{-}11\%). \textbf{e}, Top, ratio of the kinetic (\ensuremath{\tfrac12}\ensuremath{\rho}|u|\textsuperscript{2}) to the potential (p\textsuperscript{2}/2\ensuremath{\rho}c\textsuperscript{2}) energy density in the bone at \ensuremath{\theta} = 0\textdegree{}, from the peak particle velocity and peak pressure; kinetic energy in excess of the potential energy is motion without pressure, i.e. shear. Bottom, the same ratio integrated over the bone versus rotation; orange, \ensuremath{\theta} = 0\textdegree{}; blue band and dashed line, range and mean over the misaligned angles. \textbf{f}--\textbf{h}, Transmitted ex vivo bursts from one rotation series (2~mm steps; 1~mm at 0\textdegree{}). \textbf{f}, Envelope of the 0.7--1.3~MHz band at \ensuremath{\theta} = 0\textdegree{}, in dB relative to the maximum of each trace; white lines, \ensuremath{-}6~dB contour; dotted lines, the 10--50~mm range analysed. \textbf{g}, Median envelope over 10--50~mm, aligned at the leading \ensuremath{-}6~dB edge, for \ensuremath{\theta} = 0\textdegree{} and for the misaligned angles (median and interquartile range of the per-angle medians), with the 0\textdegree{} envelope plus a copy delayed by 2.5~\textmu s, the delay expected for a shear path through about 7~mm of bone. \textbf{h}, \ensuremath{-}6~dB burst width, median and interquartile range over positions; black, \ensuremath{\theta} = 0\textdegree{}, grey band its interquartile range; orange region, widths 2.5~\textmu s or more above the 0\textdegree{} value, where a shear path would place the bursts.}
\label{fig:6}
\end{figure*}

\textbf{Fluid versus elastic models.} Fluid solvers neglect shear, so we compared 2D fluid and elastic simulations of the same skull slice (Fig.~\ref{fig:6}a--c). Including shear lowered the peak pressure inside the bone (0.65 vs 1.12~MPa on axis at \ensuremath{\theta} = 0\textdegree{}) and at the focus (0.51 vs 0.57~MPa; Fig.~\ref{fig:6}d). In the elastic runs the kinetic energy in the bone was 2.7--2.9 times the energy derived from pressure at every angle, consistent with substantial shear motion. This ratio was lowest at alignment (2.74 at 0\textdegree{}) and 3--5\% higher at every other angle between \ensuremath{-}6\textdegree{} and 6\textdegree{}, without growing with the size of the rotation (Fig.~\ref{fig:6}e); the geometric incidence metric shows the same step. Misalignment therefore adds a small amount of shear-type motion in these 2D simulations, but no progressive, angle-dependent trapping of shear energy.

\textbf{Parametric amplification via velocity mismatch.} As derived in Section~\ref{sec:5.2}, the efficiency of nonlinear parametric interaction is theoretically proportional to \textit{c}\textsuperscript{\ensuremath{-}3}. Assuming this scaling can be extrapolated from fluids to elastodynamic modes in a solid, the slower velocity of shear waves in bone (\textit{c\textsubscript{S}} \ensuremath{\approx} 1400~m/s) compared to longitudinal waves (\textit{c\textsubscript{L}} \ensuremath{\approx} 2800~m/s) implies that a unit of shear energy could be approximately eight times more efficient at generating nonlinear byproducts. By potentially converting fast-propagating waves into slow-propagating ones, the misaligned skull could effectively force acoustic energy to linger inside the highly nonlinear diploe layer for twice the duration, acting as a more efficient source.

This shear-amplification picture is speculative. No existing solver couples 3D elastodynamics with dual-frequency nonlinear mixing. Rapid shear attenuation in bone may quench the effect before it accumulates. An underestimated \textit{B}/\textit{A} or fixture slop remain simpler, equally plausible explanations. The practical conclusion is unaffected: the PA dip at alignment is repeatable in both computation and experiment, regardless of which mechanism dominates the steepness.

\textbf{Waveform duration.} A parametric signal generated through a shear path would arrive later than one generated by longitudinal waves; for about 7~mm of bone the extra delay is about 2.5~\textmu s. We therefore measured the \ensuremath{-}6~dB width of the 1~MHz envelope of the transmitted bursts at each rotation angle of one ex vivo series (Fig.~\ref{fig:6}f--h). The median width was 27.1~\textmu s at 0\textdegree{} and 26.5--31.6~\textmu s at the 16 other angles, with no systematic dependence on |\ensuremath{\theta}|; only at \ensuremath{-}4\textdegree{} did it exceed the 0\textdegree{} value by more than 2.5~\textmu s, and the aligned and misaligned envelopes ended at the same time (Fig.~\ref{fig:6}g,h). The bursts therefore did not lengthen with misalignment as a shear path would require.

\section{Discussion}\label{sec:3}

To comply with the strict targeting requirements in the brain, where mistargeting can introduce severe safety risks or when accurately targeting specific brain regions or neuronal circuits is of the essence \cite{gavrilov1996,airan2017}, a millimeter (i.e., sub-wavelength) targeting accuracy is mandatory. To achieve this level of accuracy, our investigations addressed the long-standing challenge of skull-compensating lens registration by uncovering the close relationship between skull nonlinearity and the aberrations caused by misregistration. This observation allowed us to leverage low frequency acoustic feedback generated by the nonlinear mixing of high-frequency beams as they propagate through the highly non-linear skull to precisely align with the patient's anatomy. This precision extends well beyond what current neuronavigational approaches can achieve, which is at best 2~mm\cite{choi2024,chen2021,wei2013,pouliopoulos2020}. The proposed PA feedback method complements macroscopic neuronavigation by providing the critical sub-wavelength alignment required for high-frequency TUS, and our simulations demonstrate that it is robust to rotational (Fig.~\ref{fig:3}) misregistration, including the translations that accompany rotation about the fixture pivot (Table~\ref{tab:1}). Regarding the clinically acceptable margin of error, our experimental data demonstrated that a registration tolerance of \ensuremath{\pm}1\textdegree{} was acceptable for the specific skull segment tested-maintaining targeting accuracy and focal pressure within safe therapeutic margins. However, human skulls exhibit high inter-patient variability in thickness, porosity, and internal composition. Consequently, establishing a universal clinical tolerance for registration will require future studies across a diverse dataset of varying skull geometries, though the core mechanism of tracking the PA signal minimum remains robust across variations.

Beyond its immediate application to skull-compensating lens registration, we envision that the PA acoustic feedback could also be utilized for noninvasive aberration correction of dense phased arrays, acting as an objective functional metric for in vivo phase and amplitude optimization of individual elements. Together, these conceptual contributions and advancements support the design of simple, economical, and high-performance ultrasound systems for high precision neurointerventions. Such systems may support daily or weekly treatments in outpatient or limited-resource settings without compromising performance, thereby enabling the effective translation and broad clinical dissemination of this technology\cite{schoen2022,rincontorroella2022}. It may also alleviate the heavy reliance on repeated intraoperative MRI during TUS interventions, such as targeted drug delivery or liquid biopsy, that can significantly complicate or delay their implementation (e.g., the average wait time for an MRI appointment can be several months\cite{hofmann2023}).

While our experimental and numerical results both indicated a substantial drop in the PA signal during optimal alignment, we recognized a vital discrepancy: the experimental sensitivity curve exhibited a significantly steeper and more robust angular drop-off than scalar fluid-only simulations could predict. By integrating a full-wave elastodynamic geometric analysis, we propose a candidate mechanism: the skull may act not only as a passive aberrator but also as a geometrically triggered shear-energy trap that could amplify the parametric-array signal during misalignment. As shown by our geometric incidence analysis (Methods~\ref{sec:5.2}), rotational errors push part of the acoustic aperture beyond the longitudinal critical angle (\ensuremath{\theta}c \ensuremath{\approx} 30\textdegree{}). Our 2D elastic simulations showed only a small, step-like increase of shear-type motion on misalignment, and the ex vivo bursts did not lengthen (Fig.~\ref{fig:6}), so this mechanism is not supported by our current data.

Because shear waves in bone propagate at roughly half the velocity of longitudinal waves (c\textsubscript{S} \ensuremath{\approx} 1400~m/s vs. c\textsubscript{L} \ensuremath{\approx} 2800~m/s), they would increase the local acoustic energy density and extend the interaction time. If the inverse-cube scaling (c\textsuperscript{\ensuremath{-}3}) of the Westervelt source term (Section~\ref{sec:5.2}) extends to shear modes, slow-moving shear energy could generate the difference frequency up to about eight times more efficiently ((2800/1400)\textsuperscript{3} \ensuremath{\approx} 8) than equivalent compressional energy. This could contribute to the steep angular sensitivity observed experimentally, but it remains a hypothesis: no current solver couples 3D elastodynamics with dual-frequency nonlinear mixing. Separately, spectral analysis of the transmitted signal showed a clear difference-frequency peak and no subharmonic emission (f\textsubscript{0}/2), which argues against microbubble cavitation as the source of the nonlinearity (Supplementary Information).

If confirmed, this mode conversion would make the skull amplify alignment errors, turning geometric misregistration into a more easily detectable acoustic signal. Testing it will require frequency-resolved measurements of skull shear properties and solvers that couple elastodynamics with nonlinear mixing.

The nonlinearity parameter \ensuremath{\beta}, which fundamentally governs PA signal generation, exhibits frequency-dependent behavior that is often oversimplified in simulations\cite{panfilova2021,zhang2001}. Additionally, \ensuremath{\beta} for the skull, which is a highly porous structure, has not been characterized in the literature, suggesting that the current values used in theoretical investigations may not be optimal. The experimental uncertainties may also have contributed to this. Most notably, microscopic air bubbles trapped in skull pores may persist\cite{tang2011} despite the extended degassing we performed (see Methods). These microbubbles, which are resonant in the MHz range and exhibit extreme nonlinearity even at very low void fractions, may contribute to skull non-linearity \cite{cavaro2011,overvelde2010}. Additionally, the skull bone follows complex frequency-dependent attenuation and has high interindividual variability that is often underestimated in simulations \cite{pinton2011,pinton2012}. Together, these sources of uncertainty can lead to a higher Goldberg number (i.e., nonlinearities) and PA signal under the experimental conditions. Accurate, frequency-resolved measurements of skull \textit{B}/\textit{A} and shear parameters are now a priority. The principal validation stands: the PA signal drops at optimal alignment in both computation and experiment, confirming its value as a registration metric.

Collectively, the proposed research by introducing robust registration strategies to support the design of high-fidelity transcranial holography, is poised to support the effective translation and broad dissemination of this promising technology to the clinics. Our research also lays the groundwork for future research exploring the low frequency nonlinear acoustic feedback for the diagnosis, monitoring, and treatment of brain diseases and highlights the importance of relatively thin and highly nonlinear media to augment finite-amplitude effects.

\section{Conclusions}\label{sec:4}

To conclude, nonlinear wave propagation through the skull provides a registration signal for holographic lenses. The key findings are:

\begin{itemize}
\item \textbf{PA feedback as a registration metric. }The difference-frequency signal generated inside the skull drops to a minimum when the lens is correctly aligned. This dip was observed in both fluid simulations and ex vivo experiment.
\item \textbf{Energy trapping and the shear hypothesis. }Misalignment traps primary-beam energy in the high-\ensuremath{\beta} bone, increasing the integrated squared primary pressure in the bone by about 9\% (5\% per bone voxel). A secondary hypothesis, that mode-converted shear waves amplify the effect via a c\textsuperscript{\ensuremath{-}3} efficiency scaling, was not supported: the 2D elastic simulations showed only a 3--5\% increase in shear-type motion on misalignment, and the transmitted bursts did not lengthen; an underestimated B/A and fixture error are equally plausible explanations for the steep experimental roll-off.
\item \textbf{Intrinsic skull nonlinearity. }Distance-dependent measurements exclude hydrophone pseudo-sound, and the absence of subharmonic spectral lines argues against microbubble cavitation. The registration signal is consistent with cumulative quadratic nonlinearity of the bone matrix.
\item \textbf{Spatial robustness (simulation). }The dip at 0\textdegree{} was present at 88\% of receiver positions within a 65 \ensuremath{\times} 60~mm region beyond the skull, and at 98\% with a 20~mm receiver aperture.
\end{itemize}

\section{Methods}\label{sec:5}

\subsection{Hologram Registration Approach}\label{sec:5.1}

\subsubsection{Mechanism of Lens Registration}\label{sec:5.1.1}

To establish the theoretical basis for our registration strategy, we modeled the interaction of finite amplitude ultrasound waves within a medium of spatially varying nonlinearity. We utilize the Westervelt equation to model the generation of the difference frequency component, where the finite-amplitude primary waves act as a volumetric driving source.

\textbf{The nonlinear source term.} Under the quasilinear approximation, the secondary difference frequency field \ensuremath{p_{\Delta f}}(r, \textit{t}) is driven by a virtual volumetric source density, \ensuremath{S_{\mathrm{NL}}}:

\begin{equation}
\nabla^{2}p_{\Delta f}-\frac{1}{c_0^{2}}\frac{\partial^{2}p_{\Delta f}}{\partial t^{2}}=-\underbrace{\frac{\beta(r)}{\rho_0c_0^{4}}\frac{\partial^{2}}{\partial t^{2}}\left\langle p_{\mathrm{primary}}^{2}\right\rangle}_{S_{\mathrm{NL}}(r,t)}
\label{eq:1}
\end{equation}

\noindent Where:

\begin{itemize}
\item \ensuremath{\beta}(r) is the spatially dependent coefficient of nonlinearity.
\item \ensuremath{\langle}\ensuremath{p_{\mathrm{primary}}}\textsuperscript{2}\ensuremath{\rangle} is the envelope of the squared primary pressure field.
\item \ensuremath{S_{\mathrm{NL}}} represents the local strength of nonlinear generation.
\end{itemize}

The amplitude of the parametric signal is proportional to the volume integral of the source term over the interaction domain \textit{V}:

\begin{equation}
P_{\Delta f}(r_{\mathrm{obs}})\propto\int_{V}\frac{\beta(r)}{\rho_0c_0^{4}}\left|p_{\mathrm{primary}}(r)\right|^{2}G(r,r_{\mathrm{obs}})\,\mathrm{d}V
\label{eq:2}
\end{equation}

\noindent \textbf{Material contrast.} To determine the sensitivity of this method, we analyzed the significant contrast in the nonlinearity parameter \ensuremath{\beta} between the skull bone and the surrounding soft tissue.

\begin{itemize}
\item \textbf{Soft Tissue / Water:} \ensuremath{\beta} \ensuremath{\approx} 3.5--4.5.
\item \textbf{Skull Bone:} \ensuremath{\beta} \ensuremath{\approx} 188 (derived from \textit{B}/\textit{A} \ensuremath{\approx} 374).
\end{itemize}

Because \ensuremath{\beta_{\mathrm{skull}}} \ensuremath{\gg} \ensuremath{\beta_{\mathrm{tissue}}}, the integral in equation~(\ref{eq:2}) is primarily dominated by the volume of the skull illuminated by high-intensity ultrasound.

Two limiting cases bound the expected PA emission:

\begin{itemize}
\item \textbf{Misaligned:} In this state, the lens no longer corrects the skull's aberration profile. Primary beam energy scatters and reverberates inside the porous bone (high \ensuremath{\beta}). The product \ensuremath{\beta}(r)\ensuremath{\cdot}|\ensuremath{p_{\mathrm{primary}}}|\textsuperscript{2} is large throughout the bone volume, so the skull radiates a strong PA signal.
\item \textbf{Aligned:} Phase aberration due to the skull are compensated; constructive interference occurs beyond the skull. Primary energy propagates through the bone quickly and focuses in low- \ensuremath{\beta} brain tissue. The overlap integral (equation~(\ref{eq:2})) is minimized under this condition.
\end{itemize}

\subsubsection{Leveraging Nonlinear Parametric Array Effect}\label{sec:5.1.2}

The parametric array effect is a phenomenon in which nonlinear interaction between two high frequency sound beams \ensuremath{f_{1}} and \ensuremath{f_{2}} generates sum and difference frequencies \ensuremath{f_{\pm }} = \ensuremath{f_{1}}\ensuremath{\pm}\ensuremath{f_{2}} under finite-amplitude (i.e., high-intensity) wave propagation. Here, the parametric difference frequency \ensuremath{f_{- }}is written as \ensuremath{\Delta}\textit{f}. This occurs when collimated sound beams with slightly different frequencies interact nonlinearly \cite{ingard1955,westervelt1957}, the difference frequency generation-termed the parametric array effect enables focusing of low frequency pressure with sub-wavelength precision, which has been explored for the inference of tissue stiffness in elastography\cite{fatemi1998,konofagou2001,karjalainen1999}. The parametric pressure at the difference frequency can be expressed via the modulating envelope \textit{E}(\textit{t}) = 2cos(\ensuremath{\Delta}\textit{\ensuremath{\omega}t}/2) \cite{pompei2002,berktay1965} as

\begin{equation}
p_{\Delta f}\propto\beta p_0^{2}\frac{\mathrm{d}^{2}E^{2}}{\mathrm{d}t^{2}}
\label{eq:3}
\end{equation}

\noindent where \ensuremath{p_{0}} and \ensuremath{p_{\Delta f}} are the primary and difference frequency waves, and \ensuremath{\beta} is the nonlinearity parameter and \ensuremath{\omega} = 2\textit{\ensuremath{\pi}f} is the angular frequency. This relation holds for both planar and focused sources, with the difference frequency amplitude directly proportional to the primary amplitude and medium nonlinearity. We utilize the proportionality \ensuremath{p_{\Delta f}} \ensuremath{\propto} \textit{\ensuremath{\beta}p}\textsubscript{0}\textsuperscript{2} to detect lens misregistration relative to the skull. By driving the transducer with two primary waves around a high frequency (e.g., 1~MHz) differing by a low difference frequency (e.g., 100~kHz), we focus the difference frequency in the brain within the weakly nonlinear regime. The nonlinearity parameter \ensuremath{\beta} for high-density trabecular bone using B/A = 374 \cite{renaud2008} and $\beta =1+\frac{B}{2A}$, (\ensuremath{\beta_{\mathrm{bone}}} = 188) is much higher than that for coupling media like water (\ensuremath{\beta_{\mathrm{water}}} = 3.5) and brain (\ensuremath{\beta_{\mathrm{brain}}} = 4.45). Thus, variations in the parametric pressure are sensitive to skull-lens misregistration.

\subsubsection{Practical Implementation: Double-Layer Propagation}\label{sec:5.1.3}

\textbf{Frequency-dependent transmission.} For external detection, the PA signal must cross the distal skull layer. The primary beam (\ensuremath{f_{0}} \ensuremath{\approx} 1~MHz) after the focal plane attenuates heavily on this second pass (\ensuremath{\alpha} \ensuremath{\propto} \textit{f\textsuperscript{b}}), but the 100~kHz difference frequency crosses the distal skull with negligible loss.

\begin{itemize}
\item At \ensuremath{f_{0}} = 1~MHz, attenuation is high (\ensuremath{\approx}8~dB/cm in our skull model, which uses a power law with exponent y = 2). The primary beam is strongly attenuated by the exit layer.
\item At \ensuremath{\Delta}\textit{f} = 100~kHz, attenuation is negligible (\ensuremath{\approx}0.08~dB/cm).
\end{itemize}

Therefore, the PA signal generated at the entry layer propagates through the brain and exits the skull layer with minimal energy loss.

\subsection{Calculation of Elastodynamic Energy Partitioning and Parametric Amplification}\label{sec:5.2}

\textbf{Geometric Incidence and Critical Angle.} The threshold for longitudinal wave transmission at the fluid-bone interface is analytically quantified using Snell's law for elastic media. Assuming that the speed of sound in the water coupling medium is \textit{c\textsubscript{w}} \ensuremath{\approx} 1480~m/s, and the longitudinal speed of sound in the cortical bone is \textit{c\textsubscript{b}} \ensuremath{\approx} 2800~m/s, the critical angle \ensuremath{\theta}\textsubscript{\textit{c}} is

\begin{equation}
\theta_c=\arcsin\left(\frac{c_w}{c_b}\right)\approx\arcsin(0.528)\approx31.9^{\circ}
\label{eq:4}
\end{equation}

\noindent For the shear-risk metric this value was rounded to 30\textdegree{}. The metric is the percentage of vertices of the skull surface mesh (all surfaces of the binarized CT volume, including internal trabecular surfaces) that lie inside the 60~mm beam cylinder and whose outward normal makes an angle above 30\textdegree{} with the beam axis. It assumes plane-wave incidence along the axis, whereas the converging beam of the F\# 0.75 lens has marginal rays at about 34\textdegree{}. The 2D elastic simulations (Fig.~\ref{fig:6}) used the CT-derived compressional sound speed and density of the skull slice, with a shear speed of 1400~m/s in bone voxels (c > 1700~m/s).

\textbf{Stress Tensor and Energy Densities.} In the k-Wave elastic solver, the macroscopic scalar pressure \textit{P} is derived exclusively from the trace of the stress tensor (normal stresses):

\begin{equation}
P=-\frac{1}{2}\left(\sigma_{xx}+\sigma_{yy}\right)
\label{eq:5}
\end{equation}

\noindent To rigorously quantify the elastodynamic partitioning of acoustic energy within a solid skull matrix, the total energy density was separated into kinetic and potential components. The Kinetic Energy Density (\textit{w\textsubscript{k}}), which captures the total particle motion (including both compressional and shear contributions), is defined via the peak particle velocity vector u = (\textit{u\textsubscript{x}}, \textit{u\textsubscript{y}}):

\begin{equation}
w_k(r)=\frac{1}{2}\rho(r)\left(\left|u_x(r)\right|^{2}+\left|u_y(r)\right|^{2}\right)
\label{eq:6}
\end{equation}

\noindent where \ensuremath{\rho}(r) denotes the local mass density. The Potential Energy Density (\textit{w\textsubscript{p}}), representing the energy stored purely in volumetric compression, is approximated using the peak scalar pressure \textit{P}:

\begin{equation}
w_p(r)=\frac{1}{2}\frac{P(r)^{2}}{\rho(r)\,c_L(r)^{2}}
\label{eq:7}
\end{equation}

\noindent where \textit{c\textsubscript{L}} is the speed of longitudinal sound. Because pure shear waves are isochoric, their energy is captured almost exclusively by the kinetic term \textit{w\textsubscript{k}}. The total trapped energy, \ensuremath{E_{\mathrm{total}}}, is obtained by volume-integrating these densities over the spatial domain of the skull mask, \ensuremath{\Omega}\textsubscript{skull}:

\begin{equation}
E_{\mathrm{total}}=\sum_{r\in\Omega_{\mathrm{skull}}}\left(w_k(r)+w_p(r)\right)\Delta x\,\Delta y
\label{eq:8}
\end{equation}

\noindent \textbf{Derivation of the \textit{c}}\textbf{\textsuperscript{\ensuremath{-}3}}\textbf{ Scaling in Bone.} The difference-frequency pressure \ensuremath{p_{\mathrm{df}}} is derived from the Westervelt equation source term \textit{S} \cite{hamilton1997,blackstock2000}:

\begin{equation}
\nabla^{2}p_{\mathrm{df}}-\frac{1}{c^{2}}\frac{\partial^{2}p_{\mathrm{df}}}{\partial t^{2}}=-\underbrace{\frac{\beta}{\rho c^{4}}\frac{\partial^{2}p_{p}^{2}}{\partial t^{2}}}_{S}
\label{eq:9}
\end{equation}

\noindent The local source strength scales as:

\begin{equation}
S\propto\frac{\beta\cdot\Delta\omega^{2}}{\rho c^{4}}
\label{eq:10}
\end{equation}

\noindent In the absorption-limited case (common in cortical bone), the total accumulated pressure \ensuremath{P_{\mathrm{df}}} is the integral of the source over the effective interaction length \ensuremath{L_{\mathrm{eff}}} = 1/\ensuremath{\alpha}:

\begin{equation}
P_{\mathrm{df}}\approx\int_{0}^{\infty}S\,e^{-\alpha z}\,\mathrm{d}z=\frac{S}{\alpha}
\label{eq:11}
\end{equation}

\noindent Assuming that the absorption coefficient \ensuremath{\alpha} in bone scales roughly with 1/\textit{c} at a fixed frequency (a simplifying assumption), the total efficiency \ensuremath{\eta} scales as:

\begin{equation}
\eta\propto\frac{1}{c^{4}}\cdot c=\frac{1}{c^{3}}
\label{eq:12}
\end{equation}

\noindent Therefore, hypothetically, for shear waves (\textit{c\textsubscript{s}} \ensuremath{\approx} 1400~m/s) vs. longitudinal waves (\textit{c\textsubscript{l}} \ensuremath{\approx} 2800~m/s):

\begin{equation}
\frac{\eta_s}{\eta_l}=\left(\frac{c_l}{c_s}\right)^{3}=2^{3}=8
\label{eq:13}
\end{equation}

\noindent This derivation suggests that if fast longitudinal waves are mode-converted into slow shear waves, an approximately 8 -fold generation efficiency boost could potentially occur. However, we present this strictly as an exploratory hypothesis to help explain possible experimental discrepancies, rather than a definitively validated physical mechanism.

\subsection{Hologram Registration Validation}\label{sec:5.3}

We employed nonlinear ultrasound simulations to simulate the impact of skull bone misregistration on the parametric field. Specifically, by solving the second-order wave equation known as Westervelt's equation, which accounts for nonlinear propagation with thermoviscous effects. For this purpose, we utilized a pseudo-spectral time domain (PSTD) approximation of Westervelt's equation implemented in the k-wave toolbox. Although the inclusion of higher-order harmonics due to nonlinearity significantly increases computational overhead, this challenge was effectively addressed through GPU-enabled parallelization. Power law absorption was modeled using the fractional Laplacian operator (\ensuremath{\nabla}\textsuperscript{2})\textit{\textsuperscript{y}}\textsuperscript{/2} \cite{chen2004}. Additionally, nonlinearity was incorporated through the following lossless discrete pressure-density relation\cite{treeby2010}:

\begin{equation}
p^{n+1}=c_0^{2}\left[\rho^{n+1}+\frac{B}{2A}\frac{1}{\rho_0}\left(\rho^{n+1}\right)^{2}\right]
\label{eq:14}
\end{equation}

\noindent where \ensuremath{\rho}\textit{\textsuperscript{n}} = \ensuremath{\rho}(\textit{r}, \textit{n}\ensuremath{\Delta}\textit{t}). The nonlinearity parameter \ensuremath{\beta} was set to 3.6 for water (B/A = 5.2) and 188 for skull bone (B/A = 374), unless stated otherwise. The simulations were performed with a spatial grid size of 250~\textmu m (six points per wavelength at 1~MHz in water) and a time step of 20 ns (CFL \ensuremath{\approx} 0.12 in water and \ensuremath{\approx} 0.25 at the highest skull sound speed). The objective was to measure the difference frequency field near the focal region under conditions of misregistration between the skull bone and the lens. Skull misregistration was induced by rotating the skull in the transverse plane (i.e., about the \textit{z}-axis) in 1\textdegree{} steps up to \ensuremath{\pm}6\textdegree{}, and the difference frequency field was subsequently measured at 100~kHz. In the literature, the nonlinearity parameter \ensuremath{\beta} for bone, which is not well-characterized, is generally ignored, due to the high attenuation and thin cross-section of the skull bone \cite{rosnitskiy2015}. To gain further insights, simulations were conducted by varying the nonlinearity parameter \ensuremath{\beta} and the absorption coefficient \ensuremath{\alpha}. These parameters were combined using Goldberg's number \ensuremath{\Gamma}, defined as \ensuremath{\Gamma} = \ensuremath{\beta}\ensuremath{\varepsilon}k/\ensuremath{\alpha} where \textit{k} is the wave number and \ensuremath{\epsilon} is the acoustic Mach number, defined by $\epsilon =\frac{{p}_{0}}{{\rho }_{0}{c}_{0}^{2}}$. The Goldberg number is the ratio of the absorption length to the shock-formation distance; the values quoted in Section~\ref{sec:2.2} use p0 = 1~MPa, \ensuremath{\rho}0 = 1000 kg/m\textsuperscript{3}, c0 = 1500~m/s and \ensuremath{\alpha} = 10~dB/cm at 1~MHz.

\subsubsection{Volumetric Analysis of Energy Trapping}\label{sec:5.3.1}

To rigorously confirm that the registration signal arose from the acoustic energy trapped within the skull, we performed a 3D volumetric analysis of the primary pressure fields obtained from the k-Wave simulations.

The nonlinear source term \ensuremath{S_{\mathrm{NL}}} is proportional to the square of the primary pressure inside the bone (\ensuremath{S_{\mathrm{NL}}} \ensuremath{\propto} \ensuremath{\beta_{\mathrm{skull}}}|\ensuremath{p_{\mathrm{primary}}}|\textsuperscript{2}). We defined the skull interaction volume via the acoustic impedance map (\textit{c} > 1700~m/s) and calculated the Integrated Source Potential (\ensuremath{E_{\mathrm{trapped}}}) specifically within the bone matrix as follows:

\begin{equation}
E_{\mathrm{trapped}}=\int_{V_{\mathrm{skull}}}\left|p_{\mathrm{primary}}(r)\right|^{2}\mathrm{d}V
\label{eq:15}
\end{equation}

\noindent The integration was restricted to axial positions of 5 to 35~mm, which contain the skull and exclude the 3D-printed lens (whose sound speed also exceeds 1700~m/s). Because rotation changes the bone volume within this window (1,348,287 vs. 1,400,462 voxels at 0\textdegree{} and 2\textdegree{}), we also report the mean squared pressure per bone voxel, which increased by 5.2\%. Over the whole grid, including the lens, the integrated value increased by 2.8\%.

\subsection{Robustness and Artifact Analysis Methodology}\label{sec:5.4}

\textbf{Robustness to Hydrophone Placement:} To determine whether the registration metric is robust to the positioning of the receiving hydrophone, a requirement for clinical translation, we evaluated its spatial invariance. We simulated a receiver (a point, or a finite aperture such as a piston hydrophone or ultrasound transducer) placed at every grid point 40--105~mm deep and within \ensuremath{\pm}30~mm laterally behind the skull (points within \ensuremath{-}20~dB of the field maximum). To quantify the benefits of using larger detectors, we performed a parameter sweep by varying the receiver aperture from 2~mm to 30~mm, modelling a finite receiver by averaging the \ensuremath{\Delta}f amplitude across its aperture.

\textbf{Pseudo-Sound Quantification:} To verify that the measured low-frequency signals originated from true parametric generation within the medium rather than from the nonlinear transfer function of the hydrophone, we quantified the contribution of pseudo-sound. Pseudo-sound arises from the nonlinearity of the hydrophone (\textit{e} \ensuremath{\approx} mp+\textit{\ensuremath{\eta}p}\textsuperscript{2}). We took advantage of near-field measurements taken at the focus, 60~mm from the focus, and 120~mm from the focus to quantify and isolate pseudo-sounds based on how true difference-frequency waves scale differently with range.

\textbf{Intrinsic Skull Nonlinearity vs. Microbubble Artifacts:} To investigate whether the observed Difference Frequency (DF) stemmed from the intrinsic classical cumulative nonlinearity of the bone matrix rather than artifactual resonant bubble nonlinearity from trapped gas, we analyzed the spectral content predicted by two competing models of nonlinearity. Degassed human skull samples were submerged in a water tank. A bichromatic excitation pulse (\ensuremath{f_{1}}, \ensuremath{f_{2}} \ensuremath{\approx} 1~MHz, \ensuremath{\Delta}\textit{f} \ensuremath{\approx} 100~kHz) was transmitted through the skull, and the resulting acoustic field was measured using a hydrophone along the propagation axis (\textit{z}). A system governed by classical modular nonlinearity produces only integer linear combinations of input frequencies, whereas microbubble artifacts are characterized by bifurcation and the emission of subharmonics (\ensuremath{f_{0}}/2).

\textbf{Simulation of Intracranial Trapped Gas:} To evaluate the robustness of the proposed acoustic lens system against realistic postoperative conditions (e.g., pneumocephalus), we simulated the effect of trapped gas within the porous structure of the skull. The simulation utilized the kwave toolbox to solve nonlinear coupled wave equations. A morphological void-filling algorithm assigned the acoustic properties of air to a specified volume fraction (\ensuremath{\phi_{\mathrm{gas}}}) of the voids, and the source was driven by dual-frequency excitation to generate a difference frequency (\ensuremath{\Delta}\textit{f} = 100~kHz) at the target depth.

\subsection{Trans-Skull Experiments}\label{sec:5.5}

The ex-vivo trans-skull experiments were performed in a degassed deionized water tank setup. A 60~mm diameter piston transducer (Precision Acoustics, Dorchester, UK) was coupled to a 3D printed skull-compensating holographic lens. The transducer was driven by a function generator (HP Agilent Keysight, 33511B) and amplified by a power amplifier (E\&I, Rochester, NY, USA). To generate the parametric array signal, the transducer was excited with a bi-frequency input signal containing 0.95~MHz and 1.05~MHz. The transducer and lens assembly were then attached to the parietal region of an overnight degassed (approximately 12h) skull cap (Skulls Unlimited, Oklahoma City, OK, USA). The focal field of the lens was scanned with a calibrated 2~mm needle hydrophone (Precision Acoustics, Dorchester, UK) mounted on a three-axis positioning system (Velmex, Bloomfield, NY, USA) and was recorded by a digital oscilloscope (Pico Technologies, St Neots, UK).

\section*{Data availability}
All data needed to evaluate the conclusions of this study are present in the paper.

\section*{Code availability}
The simulation and analysis scripts (MATLAB k-Wave and Python) will be made available upon publication.

\section*{Acknowledgements}
We thank Dr. Stas Emelianov for providing the E\&I high-power amplifier. We also thank Drs. S. Schoen Jr. and L. Degertekin, as well as the LEAP HI team, for their valuable input, especially during the early phase of this research. \\ This study was supported by National Science Foundation (Division of Civil, Mechanical and Manufacturing Innovation, Leading Engineering for America's Prosperity, Health, and Infrastructure) Grants 1933158 and 1830577 (P.D., C.A.). \\

\section*{Author contributions}
P.D. and C.A. designed the research. P.D. performed the research, analyzed the data and wrote the paper. C.A. supervised the project and revised the manuscript. \\

\section*{Competing interests}
The authors declare no competing interests.

\section*{Supplementary information}
Supplementary Sections S1--S6, Figs.~S1--S7 and equations~(S1)--(S9) follow the references of the main text.

\clearpage
\onecolumn
\setcounter{section}{0}\setcounter{figure}{0}\setcounter{equation}{0}\setcounter{table}{0}
\renewcommand{\thesection}{S\arabic{section}}
\renewcommand{\thefigure}{S\arabic{figure}}
\renewcommand{\theequation}{S\arabic{equation}}
\renewcommand{\thetable}{S\arabic{table}}
\makeatletter\setlength{\@fptop}{0pt}\setlength{\@fpsep}{14pt}\setlength{\@fpbot}{0pt plus 1fil}\makeatother
\fancyhead[R]{\footnotesize\sffamily\color{accent}Supplementary Information\enspace$\cdot$\enspace Hologram registration by parametric feedback}

\thispagestyle{firstpage}
\begin{flushleft}
{\sffamily\footnotesize\color{accent}\textbf{SUPPLEMENTARY INFORMATION}\par}
\vspace{6pt}
{\sffamily\bfseries\fontsize{17}{21}\selectfont Sub-wavelength registration of skull-compensating acoustic holograms via nonlinear parametric array feedback\par}
\vspace{9pt}
{\large Pradosh P. Dash$^{1}$ and Costas D. Arvanitis$^{1,2}$\par}
\vspace{4pt}
{\footnotesize $^{1}$George W. Woodruff School of Mechanical Engineering, Georgia Institute of Technology, Atlanta, GA 30332, USA.
$^{2}$Wallace H. Coulter Department of Biomedical Engineering, Georgia Institute of Technology and Emory University, Atlanta, GA
30332, USA.\par}
\vspace{6pt}
{\color{accent}\rule{\textwidth}{1.2pt}}
\end{flushleft}

\section{Process flow chart for trans-skull hologram registration in a clinical setting}\label{sec:S1}

\begin{figure}[h!]
\centering
\includegraphics[width=\textwidth,height=0.80\textheight,keepaspectratio]{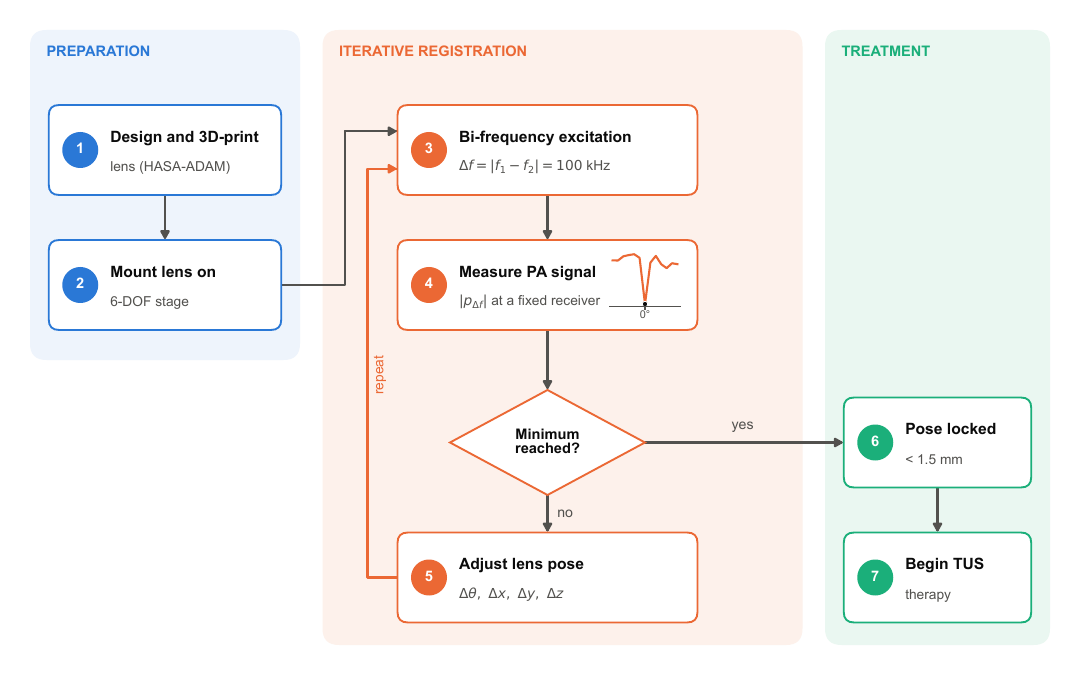}
\caption{\textbf{Process flow chart for trans-skull hologram registration in a clinical setting.} We mount a 3D-printed acoustic lens (fabricated via HASA-ADAM) on a 6-DOF stage and align it for transcranial ultrasound (TUS) therapy using an iterative process. At each iteration, we drive the transducer with bi-frequency excitation (\ensuremath{\Delta}f = |f1 \ensuremath{-} f2| = 100~kHz) and record the PA signal magnitude |p\ensuremath{\Delta}f| at a fixed receiver. We seek a minimum at \ensuremath{\theta} = 0\textdegree{}; the curve in step 4 is the simulated rotation curve of the \ensuremath{\Delta}f amplitude (B/A = 374, as in Fig.~5b). If the loop has not converged, we apply positional and angular corrections (\ensuremath{\Delta}\ensuremath{\theta}, \ensuremath{\Delta}x, \ensuremath{\Delta}y, \ensuremath{\Delta}z) and repeat. Once converged, we lock the lens position and initiate TUS therapy, achieving a targeting accuracy of < 1.5~mm.}
\label{fig:S1}
\end{figure}

\section{Robustness to Receiver Alignment for Hologram Registration using PA Signal}\label{sec:S2}

The directivity of the parametric source is set by the size of the nonlinear interaction region relative to the difference-frequency wavelength.

\begin{itemize}
\item \textbf{Primary Beam (\ensuremath{\lambda}}\textbf{ \ensuremath{\approx} 1.5~mm):} Highly directional; requires precise targeting.
\item \textbf{Parametric signal (\ensuremath{\lambda}\ensuremath{\Delta}f \ensuremath{\approx} 15~mm):} the skull (L \ensuremath{\approx} 7~mm) is thinner than \ensuremath{\lambda}\ensuremath{\Delta}f, so the parametric radiation is less directional than the primary beam.
\end{itemize}

A receiver therefore does not need to be aligned with the primary beam. Fig.~\ref{fig:S2} quantifies how the alignment dip varies with receiver position and size.

\subsection{Robustness to Hydrophone Placement: Spatial Invariance Analysis}\label{sec:S2.1}

A critical requirement for clinical translation is the insensitivity of the registration metric to the precise positioning of the receiving hydrophone. We tested this by simulating a receiver (a point, or a finite aperture such as a piston hydrophone or ultrasound transducer) placed at every grid point 40--105~mm deep and within \ensuremath{\pm}30~mm laterally behind the skull (62,400 points within \ensuremath{-}20~dB of the field maximum).

\subsubsection{Impact of Receiver Aperture Size}\label{sec:S2.1.1}

To quantify the benefit of using larger detectors, we performed a parameter sweep by varying the receiver aperture from 2~mm to 30~mm.

\begin{figure}[tbp]
\centering
\includegraphics[width=\textwidth,height=0.80\textheight,keepaspectratio]{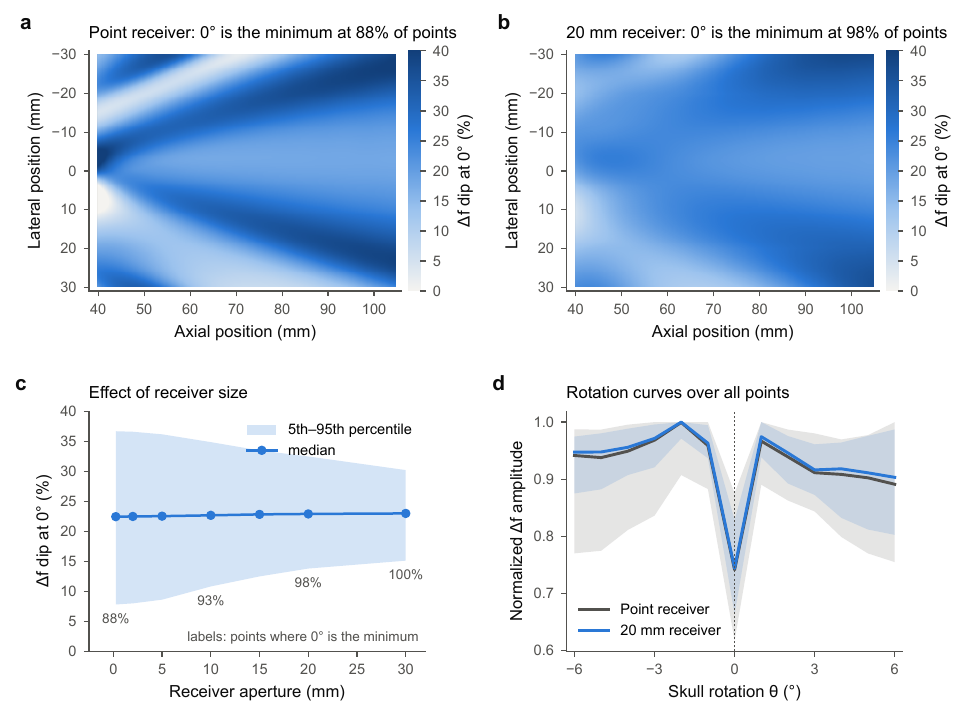}
\caption{\textbf{Robustness of the \ensuremath{\Delta}f dip to receiver position and size (3D simulation, B/A = 374).} Dip at 0\textdegree{} at every grid point 40--105~mm deep and within \ensuremath{\pm}30~mm laterally for (a) a point receiver and (b) a 20~mm receiver (\ensuremath{\Delta}f amplitude averaged across the aperture); titles give the percentage of points at which 0\textdegree{} is the minimum of the rotation curve. \textbf{c}, Median and 5th--95th percentile of the dip across all points versus receiver aperture; labels give the percentage of points at which 0\textdegree{} is the minimum. \textbf{d}, Median normalized \ensuremath{\Delta}f amplitude versus rotation (bands: 5th--95th percentile across points). Dip = 1 \ensuremath{-} A(0\textdegree{})/mean[A(\ensuremath{\pm}1\textdegree{})], with A normalized per point over \ensuremath{\theta} = \ensuremath{-}6\textdegree{} to 6\textdegree{}.}
\label{fig:S2}
\end{figure}

Increasing the receiver aperture did not change the median dip (22--23\%) but reduced its spread across positions (Fig.~\ref{fig:S2}c): the 5th-percentile dip rose from 8\% for a point receiver to 14\% for a 20~mm aperture and 15\% for a 30~mm aperture, and the fraction of positions at which 0\textdegree{} is the minimum rose from 88\% to 98\% and 100\%.

\begin{itemize}
\item A finite aperture acts as a spatial low-pass filter that averages over diffraction nulls (Fig.~\ref{fig:S2}a,b).
\item A receiver of about 20~mm, comparable to a standard clinical transducer, is therefore sufficient for robust detection in this geometry.
\end{itemize}

\section{Pseudo-Sound in Parametric Arrays}\label{sec:S3}

Pseudo-sound\cite{si-song2021,si-moffett} is a spurious low-frequency signal that appears in parametric arrays but does not originate in the acoustic medium itself. Instead, it arises from nonlinearities in the hydrophone and/or receiving electronics. When two primary high-frequency signals, \ensuremath{f_{1}} and \ensuremath{f_{2}}, are transmitted, the acoustic pressure can be modeled as

\begin{equation}
p=P_0\cos\left(\frac{\Delta\omega\,t}{2}\right)\cos\left(\omega_0t\right),
\label{eq:S1}
\end{equation}

\noindent where \ensuremath{\omega_{0}} is the carrier frequency and \ensuremath{\Delta}\ensuremath{\omega} = \ensuremath{\omega_{1}}\ensuremath{-}\ensuremath{\omega_{2}} is the difference frequency. Although a genuine difference-frequency wave is generated in the medium through the nonlinear interaction of the primary waves, an additional low-frequency signal can appear owing to the hydrophone's nonlinear transfer function.

\subsection{Nonlinear Transfer Function of the Hydrophone}\label{sec:S3.1}

A common way to express the hydrophone's approximate transfer function is as follows:

\begin{equation}
e\approx mp+\eta p^{2},
\label{eq:S2}
\end{equation}

\noindent where \textit{p} is the incident pressure, \textit{m} is the linear sensitivity, and \ensuremath{\eta} captures the nonlinear behavior of the hydrophone. Because \textit{p} contains both carrier and difference-frequency components, the \textit{p}\textsuperscript{2} term can lead to unintended low-frequency outputs, often referred to as pseudo-sound. When expanded, the square-law term produces giving rise to both DC offset and low-frequency signals at or near \ensuremath{\Delta}\textit{f}.

\begin{equation}
e\sim\eta p^{2}\rightarrow\eta P_0^{2}\times\left[\text{DC and combination frequencies}\right],
\label{eq:S3}
\end{equation}

\subsection{Experimental Quantification}\label{sec:S3.2}

\begin{figure}[tbp]
\centering
\includegraphics[width=\textwidth,height=0.80\textheight,keepaspectratio]{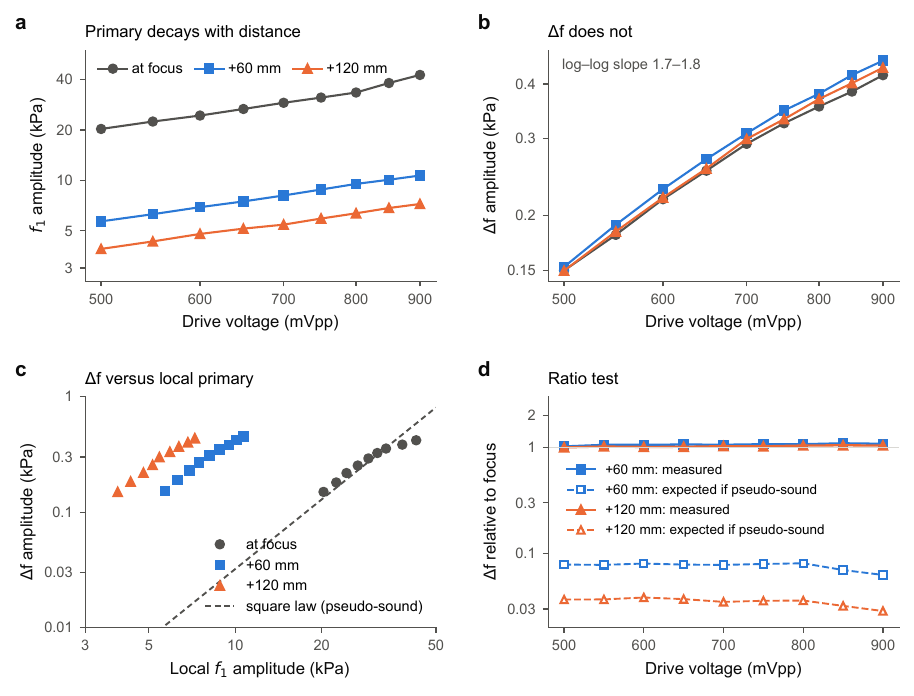}
\caption{\textbf{Pseudo-sound test in water (needle hydrophone, nine drive levels).} \textbf{a}, Primary (f\textsubscript{1} = 1.05~MHz) and (b) \ensuremath{\Delta}f (100~kHz) amplitude versus drive voltage at the focus and 60~mm and 120~mm beyond it. \textbf{c}, \ensuremath{\Delta}f amplitude versus local primary amplitude; the dashed line is a square law fitted to the focal data, the dependence expected for pseudo-sound. \textbf{d}, \ensuremath{\Delta}f amplitude at 60 and 120~mm relative to the focus (filled) and the ratio expected if the signal were pseudo-sound, (p\textsubscript{local}/p\textsubscript{focus})\textsuperscript{2} (open). The measured ratio is close to 1, 13--35 times above the pseudo-sound expectation.}
\label{fig:S3}
\end{figure}

Various strategies can quantify pseudo-sound: near-field measurements taken close to the transducer emphasize pseudo-sound because the true parametric wave has little distance to form; short-path suppression, achieved by minimizing the propagation distance in water (e.g., a small tank), diminishes the genuine difference-frequency wave and helps isolate pseudo-sound; comparing signals over multiple distances exploits the fact that real difference-frequency waves scale differently with range than pseudo-sound; and using truncators to partially block or attenuate the primary beams lowers the overall amplitude and thus square-law effects. We compared measurements in water at the focus and at 60~mm and 120~mm beyond it (Fig.~\ref{fig:S3}). Moving away from the focus lowered the local primary (f1 = 1.05~MHz) amplitude by a factor of 4--6, but the \ensuremath{\Delta}f (100~kHz) amplitude was unchanged (0.15--0.45~kPa over the drive range) and grew with drive voltage with a log--log slope of 1.7--1.8, as expected for square-law generation. Pseudo-sound from the hydrophone's square-law response scales with the local primary pressure squared and would therefore be 13--35 times weaker at 60 and 120~mm than at the focus. The measured \ensuremath{\Delta}f signal is therefore a difference-frequency wave propagating in the water, not a hydrophone artifact, at all three distances. As a precaution, the skull registration measurements were made away from the focus.

\section{Intrinsic Skull Nonlinearity: Parametric Array vs Microbubble Artifacts}\label{sec:S4}

\subsection{Approach}\label{sec:S4.1}

To validate the source of the observed Difference Frequency (DF), we analyzed the spectral content predicted by two competing models of nonlinearity: Classical Cumulative Nonlinearity (Intrinsic Bone) versus Resonant Bubble Nonlinearity (Trapped Gas).

\subsubsection{Mechanism A: Intrinsic Bone Nonlinearity}\label{sec:S4.1.1}

The propagation of high-intensity ultrasound through the skull matrix (fluid-saturated porous solid) is governed by Westervelt's equation. In the frequency domain, the nonlinearity is perturbative and can be approximated by a power series expansion of the pressure-density relation:

\begin{equation}
p_{\mathrm{out}}\approx p_{\mathrm{in}}+\beta p_{\mathrm{in}}^{2}+\ldots
\label{eq:S4}
\end{equation}

\noindent Under bichromatic excitation with frequencies \ensuremath{\omega_{1}} and \ensuremath{\omega_{2}}, the quadratic term \textit{p}\textsuperscript{2} drives nonlinearity. Substituting the input signal \textit{A}cos(\ensuremath{\omega_{1}}\textit{t})+\textit{B}cos(\ensuremath{\omega_{2}}\textit{t}) into the quadratic term yields

\begin{equation}
\begin{aligned}\left(\cos\omega_1t+\cos\omega_2t\right)^{2}={}&\underbrace{1}_{\text{DC}}+\underbrace{\tfrac{1}{2}\cos(2\omega_1t)+\tfrac{1}{2}\cos(2\omega_2t)}_{\text{second harmonics}}\\&+\underbrace{\cos\left[(\omega_1+\omega_2)t\right]}_{\text{sum frequency}}+\underbrace{\cos\left(\left|\omega_1-\omega_2\right|t\right)}_{\text{difference frequency}}\end{aligned}
\label{eq:S5}
\end{equation}

\noindent \textbf{Conclusion A:} A system governed by classical modular nonlinearity produces only integer linear combinations of the input frequencies (\textit{n\ensuremath{\omega}}\textsubscript{1}\ensuremath{\pm}\textit{m\ensuremath{\omega}}\textsubscript{2}). It is mathematically impossible for a quadratic operator to generate fractional frequencies (subharmonics), such as \ensuremath{\omega_{1}}/2.

\subsubsection{Mechanism B: Microbubble Artifacts (Period Doubling)}\label{sec:S4.1.2}

If the skull contains trapped gas pockets due to improper degassing, the nonlinearity is governed by the Rayleigh-Plesset equation. The bubble acts as a driven oscillator with a highly nonlinear stiffness owing to the compressibility of the gas. The equation of motion for the bubble radius \textit{R}(\textit{t}) is a nonlinear differential equation:

\begin{equation}
R\ddot{R}+\frac{3}{2}\dot{R}^{2}+\cdots+\omega_0^{2}R=F_{\mathrm{drive}}(t)
\label{eq:S6}
\end{equation}

\noindent Unlike the power series, this dynamic system is capable of bifurcation. When the driving acoustic pressure exceeds the Blake threshold, the stable oscillation period \textit{T} doubles to 2\textit{T}, quadruples to 4\textit{T}, or becomes chaotic.

\begin{equation}
f_{\mathrm{emit}}=\frac{1}{2T}=\frac{f_{\mathrm{drive}}}{2}\quad\text{(subharmonic)}
\label{eq:S7}
\end{equation}

\noindent \textbf{Conclusion B:} The definitive spectral marker of microbubble contamination is the presence of subharmonics (\ensuremath{f_{0}}/2) and broadband noise (spectral hash) caused by chaotic oscillation and inertial collapse.

\subsection{Experimental Results and Discussion}\label{sec:S4.2}

\begin{figure}[tbp]
\centering
\includegraphics[width=\textwidth,height=0.80\textheight,keepaspectratio]{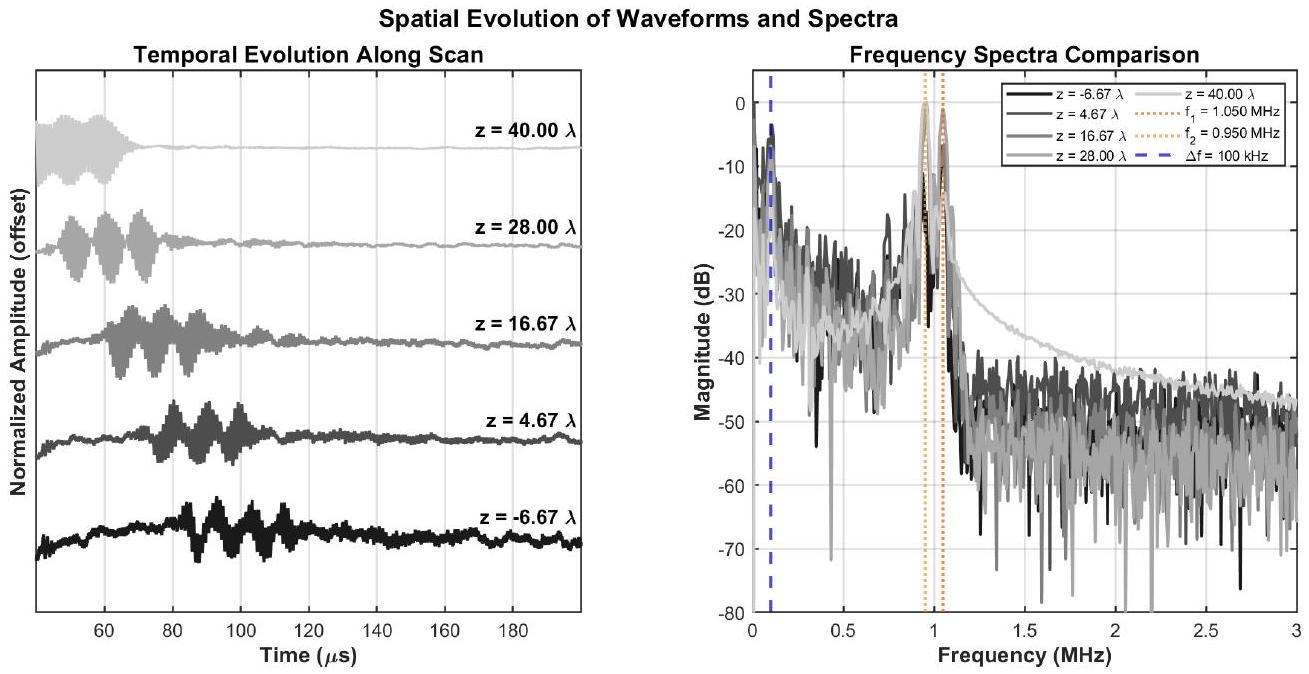}
\caption{\textbf{Time-domain waveforms and Frequency Spectra along the propagation axis.} The left column shows the raw voltage recorded by the hydrophone at increasing depths (Top to Bottom). The column on the right shows the corresponding FFT. Dashed lines indicate the Difference Frequency (\ensuremath{\Delta}\textit{f}, blue), the Primary Frequency (\ensuremath{f_{0}}, red), and the Second Harmonic (2\ensuremath{f_{0}}, green).}
\label{fig:S4}
\end{figure}

\subsubsection{Methodology}\label{sec:S4.2.1}

The degassed human skull sample was submerged in a water tank. A bichromatic excitation pulse (\ensuremath{f_{1}}, \ensuremath{f_{2}} \ensuremath{\approx} 1~MHz, \ensuremath{\Delta}\textit{f} \ensuremath{\approx} 100~kHz) was transmitted through the skull. The resulting acoustic field was measured using a hydrophone along the propagation axis (\textit{z}).

\subsubsection{Spectral Analysis}\label{sec:S4.2.2}

Fig.~\ref{fig:S4} displays the time-domain waveforms (left column) and their corresponding frequency spectra (FFT) (right column) at various depths \textit{z}. The spectral data provide evidence on the physical origin of the nonlinearity:

\begin{itemize}
\item \textbf{Confirmation of Classical Mixing:} As seen in the FFT plots (specifically positions 18 through 71), there are distinct, high-SNR peaks corresponding exactly to the mixing terms predicted by the Westervelt model: The Primary inputs (\ensuremath{f_{1}}, \ensuremath{f_{2}}) around 1~MHz and the target Difference Frequency (\ensuremath{\Delta}\textit{f}) at \ensuremath{\approx} 100~kHz.
\item \textbf{Absence of Bubble Signatures:} Crucially, we examine the spectral region corresponding to the subharmonic frequency (\ensuremath{f_{\mathrm{sub}}} \ensuremath{\approx} 500~kHz). As evidenced in the FFT plots, the noise floor in the 0.2--0.8~MHz range remained flat and unstructured. There is no detectable energy at \ensuremath{f_{0}}/2.
\end{itemize}

\subsubsection{Exclusion of Stable Cavitation Artifacts}\label{sec:S4.2.3}

While theoretical models of stable cavitation indicate that gas microbubbles can generate different frequencies via perturbative nonlinearity without triggering subharmonic emission (Regime A), this occurs only at low acoustic driving pressures. In Regime A (Stable), the scattering cross-section of a bubble is small. This is an inefficient mixer. To generate a strong, high-SNR \ensuremath{\Delta}\textit{f} peak (as shown in Fig.~\ref{fig:S4}) using only stable bubbles, we would need a massive, dense cloud that would block the primary propagation as well.

In the context of transcranial transmission, where incident pressures typically exceed the threshold for inertial cavitation and period-doubling bifurcations (often < 400~kPa for resonant bubbles), a trapped gas body is likely to be driven into the strong scattering regime (Regime B). This regime is typically characterized by the emission of subharmonics (\ensuremath{f_{0}}/2) and broadband noise. Therefore, the observation of a robust Difference Frequency signal coinciding with a complete absence of subharmonic content argues against microbubbles, in either stable or inertial regimes, as the source of the nonlinearity. The signal is therefore more likely attributable to the intrinsic, cumulative nonlinearity of the bone matrix.

\subsubsection{Exclusion of Bubble Nonlinearity via Intensity Thresholds}\label{sec:S4.2.4}

To further distinguish between the intrinsic nonlinearity of the bone matrix and artifactual nonlinearity from trapped gas microbubbles, we analyzed the signal integrity relative to the thresholds established for bubbly media by Karpov et al.\cite{si-karpov2003} demonstrated that gas-saturated layers exhibit a softening nonlinearity so profoundly that the threshold for shock wave formation is exceptionally low. In their dual-frequency simulations, they show that an incident pressure of only 50~kPa(\ensuremath{P_{\mathrm{inc}}}/\ensuremath{P_{\infty }} = 0.5) is sufficient to distort the carrier wave into a violent sawtooth profile. In contrast, our experiments utilized focal pressures of approximately 0.1~MPa(100~kPa), above the incident pressure at which Karpov et al.\cite{si-karpov2003} report shock formation. If trapped gas microbubbles were present in the diplo\"e layer, an incident pressure of 0.1~MPa would be expected to force the system into inertial cavitation. This would manifest as massive signal destruction and spectral chaos. Conclusion: Our spectral analysis (Fig.~\ref{fig:S4}) reveals high-fidelity transmission of the primary frequencies and a clean, discrete Difference Frequency (\ensuremath{\Delta}\textit{f}) peak. The ability of the medium to support 0.1~MPa propagation without degrading into the catastrophic shock regime characteristic of bubbly liquids is consistent with a propagation path free of resonant bubbles. The observed nonlinearity is therefore most likely attributable \\  to the intrinsic, cumulative parametric interaction within the solid bone matrix.

\section{Impact of Intracranial Trapped Gas on Parametric Array Generation}\label{sec:S5}

\subsection{Simulation Methodology}\label{sec:S5.1}

To evaluate the robustness of the proposed acoustic lens system against realistic post-operative or transcranial conditions, we simulated the effect of trapped gas (pneumocephalus) within the porous structure of the skull. In clinical scenarios, incomplete degassing or cavitation can result in macroscopic air pockets trapped within the diplo\"e layer or at the skull surface. Due to the acoustic impedance mismatch between fluid/bone (\textit{Z} \ensuremath{\approx} 1.5--3 MRayl) and air (\textit{Z} \ensuremath{\approx} 0.0004 MRayl), even minute gas volumes function as potent acoustic scatterers.

The simulation utilized the k-Wave toolbox \cite{si-treeby2010} to solve the nonlinear coupled wave equations. The acoustic medium was defined by a digitized skull model (\ensuremath{c_{\mathrm{skull}}} > 1700~m/s) immersed in water. To model trapped gas, a morphological void-filling algorithm identified enclosed fluid spaces within the skull geometry (excluding the initial lens interface). A stochastic model assigned the acoustic properties of air (\ensuremath{c_{\mathrm{air}}} = 343~m/s, \ensuremath{\rho_{\mathrm{air}}} = 1.2 kg/m\textsuperscript{3}, \ensuremath{\alpha_{\mathrm{air}}} = 10~dB/cm) to a specified volume fraction (\ensuremath{\phi_{\mathrm{gas}}}) of these voids.

The source was driven with a dual-frequency excitation (\ensuremath{f_{1}} = 1.05~MHz, \ensuremath{f_{2}} = 0.95~MHz) at a source pressure of 1~MPa. The setup was designed to generate a difference frequency (\ensuremath{\Delta}\textit{f} = 100~kHz) at the target depth via the parametric array effect.

\subsection{Results: The Threshold of Elimination}\label{sec:S5.2}

\begin{figure}[p]
\centering
\includegraphics[width=\textwidth,height=0.80\textheight,keepaspectratio]{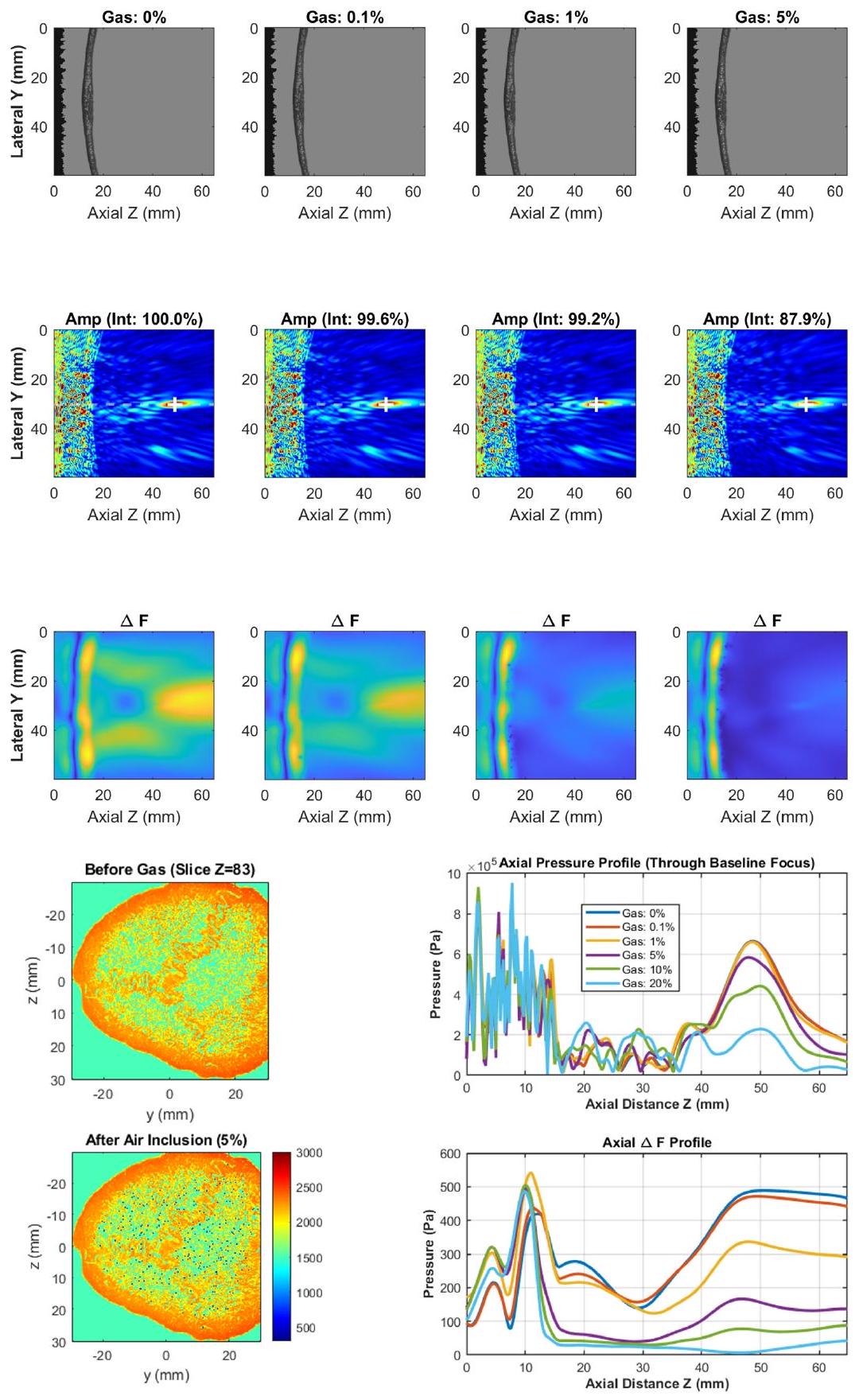}
\caption{\textbf{Effect of trapped gas on acoustic propagation.} (Top Panel) Longitudinal field maps showing the skull geometry (Row 1), Fundamental frequency amplitude (Row 2), and Difference frequency amplitude (Row 3).(Bottom Left Panel) Skull slice of speed of sound without (Row 1) and with (Row 2) gas inclusions in the microstructure (Bottom Right Panel) Axial pressure profiles extracted through the geometric focus. The top plot shows the fundamental frequency (\ensuremath{f_{1}}), while the bottom plot shows the difference frequency (\ensuremath{\Delta}\textit{f}). A critical transition occurs at \ensuremath{\phi_{\mathrm{gas}}} \ensuremath{\geq} 1\%, where the difference frequency signal is effectively extinguished.}
\label{fig:S5}
\end{figure}

The results, illustrated in Fig.~\ref{fig:S5}, demonstrate a nonlinear threshold response regarding the survival of the parametric array compared to the fundamental beam.

As shown in the axial pressure profiles (Fig.~\ref{fig:S5}, Right Panel), two distinct regimes are observed:

\begin{itemize}
\item \textbf{Weakening Regime (\ensuremath{\phi}}\textbf{\textsubscript{gas}}\textbf{ = 0.1\%):} At a gas fraction of 0.1\% (orange curve), the fundamental beam undergoes scattering but retains a focal structure. The difference frequency is similarly attenuated but remains coherent and visible.
\item \textbf{Elimination Regime (\ensuremath{\phi}}\textbf{\textsubscript{gas}}\textbf{ \ensuremath{\geq} 1\%):} A critical transition occurs at 1\% gas inclusion (yellow curve). While the fundamental beam still transmits residual energy through the skull (albeit scattered and broadened), the difference frequency signal drops to the noise floor immediately post-skull.
\end{itemize}

\subsection{Physical Interpretation: Disruption of the Nonlinear Mixing Zone}\label{sec:S5.3}

The disproportionate suppression of the difference frequency (\ensuremath{\Delta}\textit{f}) is attributed to the disruption of the nonlinear generation mechanism provided by the skull.

Contrary to acting as a passive linear barrier, the skull bone serves as an active nonlinear mixing zone. The amplitude of the parametrically generated wave, \ensuremath{P_{\Delta f}}, scales with the coefficient of nonlinearity (\ensuremath{\beta}) of the medium:

\begin{equation}
P_{\Delta f}\propto\beta\cdot P_{f1}\cdot P_{f2}^{*}
\label{eq:S8}
\end{equation}

The skull bone possesses a nonlinearity parameter significantly higher than that of water. In our model, we utilized \textit{B}/\textit{A} \ensuremath{\approx} 37.4 for the skull (resulting in \ensuremath{\beta_{\mathrm{skull}}} \ensuremath{\approx} 19.7), compared to \textit{B}/\textit{A} \ensuremath{\approx} 5.2 (\ensuremath{\beta_{\mathrm{water}}} \ensuremath{\approx} 3.6) for water. Consequently, the skull volume acts as a local amplifier for difference frequency generation, boosting the demodulation of the primary waves as they traverse the bone.

The presence of trapped gas bubbles fundamentally hampers this nonlinear engine via two mechanisms:

\begin{itemize}
\item \textbf{Decoherence of Primary Waves:} The generation of \ensuremath{\Delta}\textit{f} requires the primary waves (\ensuremath{f_{1}}, \ensuremath{f_{2}}) to propagate collinearly with high spatial coherence over an interaction length. The gas voids induce strong Rayleigh scattering, randomizing the phases of the primary waves immediately upon entry. This destroys the phase-matching condition required for the high- \ensuremath{\beta} skull medium to cumulatively build the difference frequency.
\item \textbf{Quadratic Sensitivity:} Since the difference frequency generation scales with the product of the primary pressures (equation~(\ref{eq:S8})), a linear reduction in primary amplitude due to scattering within the skull results in a quadratic reduction in the secondary source strength.
\end{itemize}

\subsection{Cavitation Threshold Analysis}\label{sec:S5.4}

It is necessary to distinguish whether the observed signal loss is driven by inertial cavitation (bubble collapse) absorbing the energy, or simply by acoustic scattering. We evaluated the cavitation threshold for the simulated parameters. The Mechanical Index (MI), a standard regulatory metric for estimating the likelihood of inertial cavitation, is defined as:

\begin{equation}
\mathrm{MI}=\frac{P_{\mathrm{neg}}\,[\mathrm{MPa}]}{\sqrt{f_c\,[\mathrm{MHz}]}}
\label{eq:S9}
\end{equation}

\noindent In our simulation, the peak source pressure is \ensuremath{P_{\mathrm{source}}} = 1.0~MPa at a center frequency of \textit{f\textsubscript{c}} = 1.0~MHz. This yields a maximum theoretical Mechanical Index of MI \ensuremath{\approx} 1.0 at the skull surface. While an MI of 1.0 approaches the cavitation threshold in bulk fluids (\textit{MI}\textsubscript{thresh} \ensuremath{\approx} 0.7--1.9), several factors indicate that cavitation is not the dominant loss mechanism here:

\begin{itemize}
\item \textbf{Attenuation in Bone:} The skull attenuates the incident wave significantly (\ensuremath{\alpha} \ensuremath{\approx} 8~dB/cm at 1~MHz in our model), reducing the local peak negative pressure at the internal void sites well below the 1~MPa surface input.
\item \textbf{Geometric Regime:} The trapped gas pockets simulated (pneumocephalus) represent macroscopic voids (> 100~\textmu m) determined by the trabecular bone structure. At 1~MHz, the resonant bubble radius is approximately 3~\textmu m. Being orders of magnitude larger than the resonance size, these gas pockets act as stiff acoustic reflectors (geometric scatterers) rather than resonating nuclei.
\end{itemize}

Therefore, we conclude that the gas inclusions act purely as passive scatterers. The elimination of the parametric array is caused by the high acoustic impedance mismatch (\ensuremath{Z_{\mathrm{skull}}}/\ensuremath{Z_{\mathrm{air}}} \ensuremath{\approx} 7500) scattering the primary energy and destroying the nonlinear coherence, rather than energy absorption via cavitation dynamics.

\section{Experimental Equipment}\label{sec:S6}

\begin{figure}[h!]
\centering
\includegraphics[width=0.47\textwidth,height=0.80\textheight,keepaspectratio]{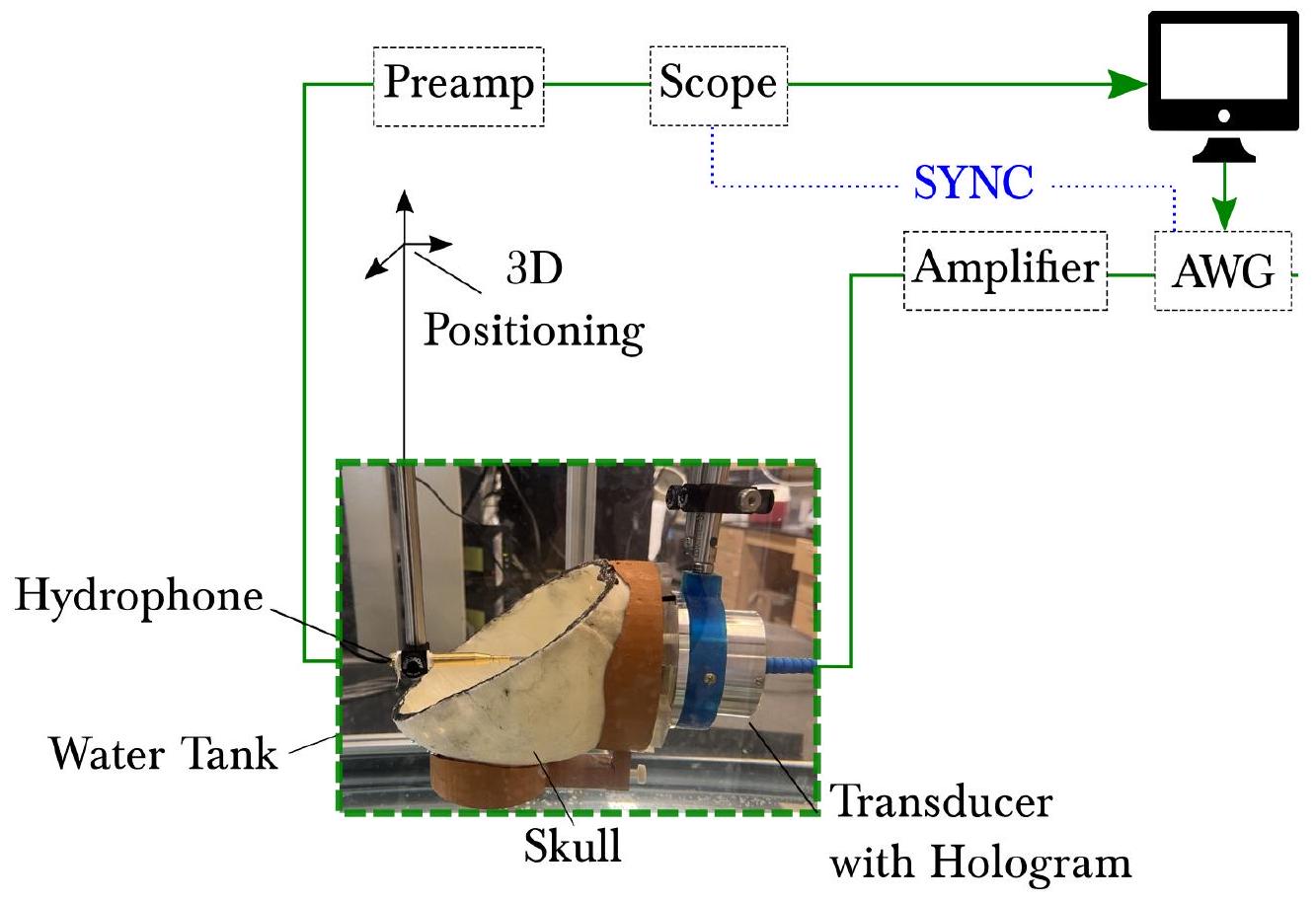}
\caption{\textbf{Experimental Setup for Hologram Validation.} An arbitrary waveform generator (AWG) generates a signal, amplified to drive a transducer with an acoustic hologram. Ultrasound waves propagate through a skull in a water tank. A hydrophone on a 3D positioning system measures the transmitted acoustic pressure field. The signal is conditioned by a preamplifier (Preamp) and digitized by an oscilloscope (Scope). A computer controls data acquisition and synchronizes (SYNC) the AWG and oscilloscope.}
\label{fig:S6}
\end{figure}

\begin{figure}[h!]
\centering
\includegraphics[width=0.47\textwidth,height=0.80\textheight,keepaspectratio]{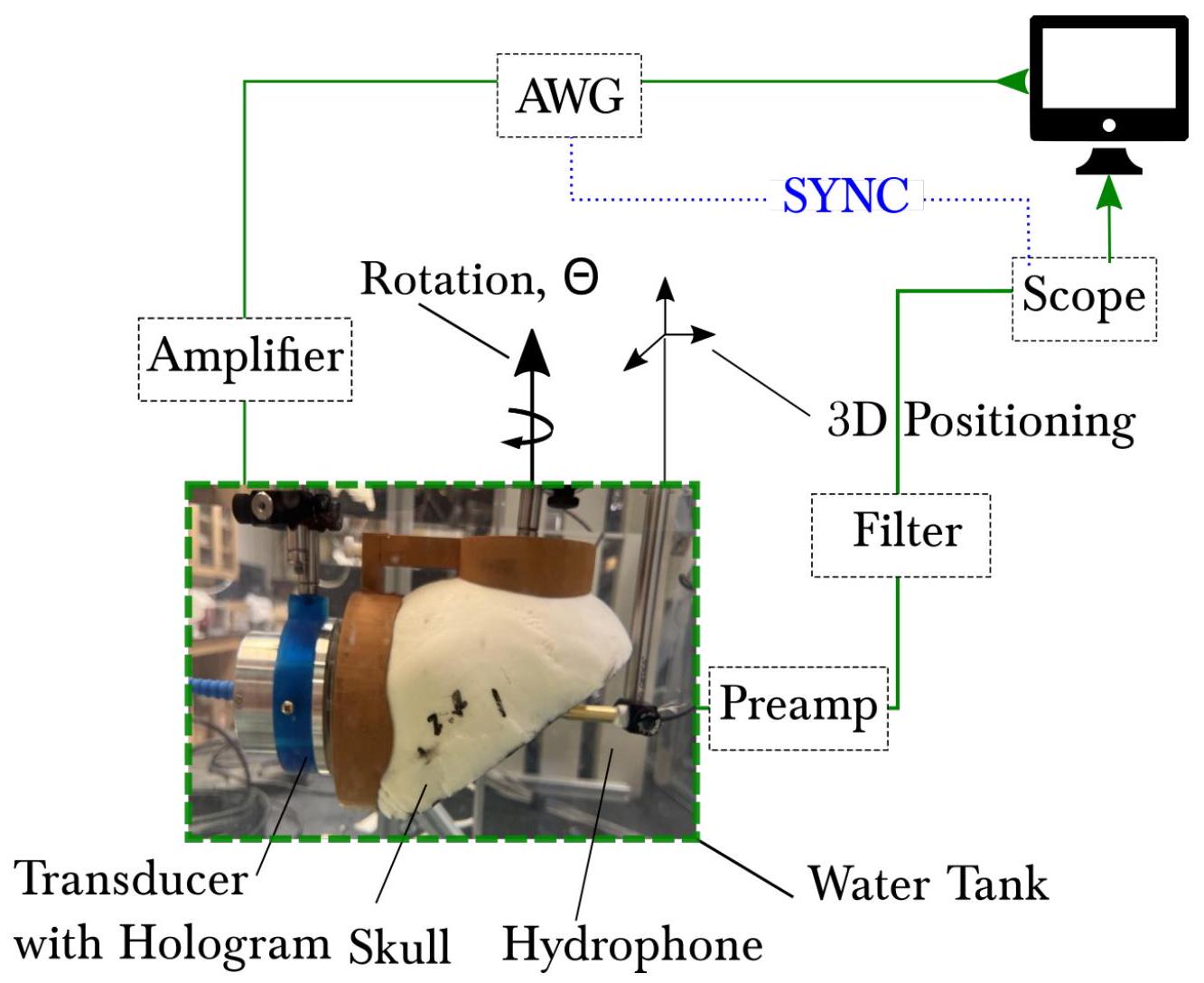}
\caption{\textbf{Experimental Setup for Hologram Registration.} We affixed the transducer, hologram, and skull assembly to a rotational axis (Rotation, \ensuremath{\Theta}) to facilitate misregistration. A distinct 3D positioning system moves the hydrophone along the z-axis. To further amplify the received PA signal, the acquisition setup routes the data through an additional 40~dB analog filter located between the preamplifier (Preamp) and the oscilloscope (Scope).}
\label{fig:S7}
\end{figure}


{\small
\begin{thebibliography}{99}
\bibitem{maimbourg2018} Maimbourg, G., Houdouin, A., Deffieux, T., Tanter, M. \& Aubry, J.-F. 3D-printed adaptive acoustic lens as a disruptive technology for transcranial ultrasound therapy using single-element transducers. \textit{Phys. Med. Biol.} \textbf{63}, 025026 (2018).
\bibitem{jimenezgambin2024} Jim\'enez-Gamb\'in, S., Bae, S., Ji, R., Tsitsos, F. \& Konofagou, E. E. Feasibility of Hologram-Assisted Bilateral Blood-Brain Barrier Opening in Non-Human Primates. \textit{IEEE Trans. Ultrason. Ferroelectr. Freq. Control} \textbf{71}, 1172--1185 (2024).
\bibitem{andres2022} Andr\'es, D., Jim\'enez, N., Benlloch, J. M. \& Camarena, F. Numerical Study of Acoustic Holograms for Deep-Brain Targeting Through the Temporal Bone Window. \textit{Ultrasound Med. Biol.} \textbf{48}, 872--886 (2022).
\bibitem{choi2024} Choi, S. W. \textit{et al.} Neuronavigation-Guided Transcranial Histotripsy (NaviTH) System. \textit{Ultrasound Med. Biol.} \textbf{50}, 1155--1166 (2024).
\bibitem{chen2021} Chen, K.-T. \textit{et al.} Neuronavigation-Guided Focused Ultrasound for Transcranial Blood-Brain Barrier Opening and Immunostimulation in Brain Tumors. \textit{Sci. Adv.} \textbf{7}, eabd0772 (2021).
\bibitem{wei2013} Wei, K.-C. \textit{et al.} Neuronavigation-Guided Focused Ultrasound-Induced Blood-Brain Barrier Opening: A Preliminary Study in Swine. \textit{Am. J. Neuroradiol.} \textbf{34}, 115--120 (2013).
\bibitem{pouliopoulos2020} Pouliopoulos, A. N. \textit{et al.} A Clinical System for Non-Invasive Blood-Brain Barrier Opening Using a Neuronavigation-Guided Single-Element Focused Ultrasound Transducer. \textit{Ultrasound Med. Biol.} \textbf{46}, 73--89 (2020).
\bibitem{westervelt1963} Westervelt, P. J. Parametric Acoustic Array. \textit{J. Acoust. Soc. Am.} \textbf{35}, 535--537 (1963).
\bibitem{dash2021} Dash, P. P. \& Arvanitis, C. Non-linearities under highly focused high-intensity ultrasound fields. \textit{J. Acoust. Soc. Am.} \textbf{150}, A125--A125 (2021).
\bibitem{hamilton1997} Hamilton, M. F. \& Blackstock, D. T. (eds) \textit{Nonlinear Acoustics} (Academic Press, San Diego, CA, 1997).
\bibitem{treeby2010} Treeby, B. E. \& Cox, B. T. k-Wave: MATLAB Toolbox for the Simulation and Reconstruction of Photoacoustic Wave Fields. \textit{J. Biomed. Opt.} \textbf{15}, 021314 (2010).
\bibitem{renaud2008} Renaud, G., Call\'e, S., Remenieras, J.-P. \& Defontaine, M. Exploration of Trabecular Bone Nonlinear Elasticity Using Time-of-Flight Modulation. \textit{IEEE Trans. Ultrason. Ferroelectr. Freq. Control} \textbf{55}, 1497--1507 (2008).
\bibitem{pinton2011} Pinton, G., Aubry, J.-F., Fink, M. \& Tanter, M. Effects of Nonlinear Ultrasound Propagation on High Intensity Brain Therapy. \textit{Med. Phys.} \textbf{38}, 1207--1216 (2011).
\bibitem{baron2009} Baron, C., Aubry, J.-F., Tanter, M., Meairs, S. \& Fink, M. Simulation of Intracranial Acoustic Fields in Clinical Trials of Sonothrombolysis. \textit{Ultrasound Med. Biol.} \textbf{35}, 1148--1158 (2009).
\bibitem{dash2025} Dash, P. P. \& Arvanitis, C. Leveraging the parametric array effect for transcranial focused ultrasound interventions. \textit{J. Acoust. Soc. Am.} \textbf{157}, A307--A307 (2025).
\bibitem{dash2025a} Dash, P. P. \& Arvanitis, C. D. Acoustic Holography in the Megahertz Frequency Range with Optimal Lens Topologies and Nonlinear Acoustic Feedback. Preprint at https://doi.org/10.48550/arXiv.2508.07103 (2025).
\bibitem{arvanitis2025} Arvanitis, C. \& Dash, P. P. Trans-skull focused ultrasound using acoustic hologram and heterogenous angular spectrum approach, and hologram registration. US Patent 12,502,558 B2 (2025).
\bibitem{dash2023} Dash, P. P. \& Arvanitis, C. Heterogenous angular spectrum approach based holograms for trans-skull focused ultrasound therapy. \textit{J. Acoust. Soc. Am.} \textbf{153}, A103--A103 (2023).
\bibitem{song2021} Song, J., Jung, D., Kim, J. S. \& Lee, J. Experimental evaluation of pseudo-sound in a parametric array. \textit{J. Acoust. Soc. Am.} \textbf{150}, 3787--3796 (2021).
\bibitem{gavrilov1996} Gavrilov, L. R., Tsirulnikov, E. M. \& Davies, I. ab I. Application of focused ultrasound for the stimulation of neural structures. \textit{Ultrasound Med. Biol.} \textbf{22}, 179--192 (1996).
\bibitem{airan2017} Airan, R. Neuromodulation with Nanoparticles. \textit{Science} \textbf{357}, 465 (2017).
\bibitem{schoen2022} Schoen, S. \textit{et al.} Towards Controlled Drug Delivery in Brain Tumors with Microbubble-Enhanced Focused Ultrasound. \textit{Adv. Drug Deliv. Rev.} \textbf{180}, 114043 (2022).
\bibitem{rincontorroella2022} Rincon-Torroella, J., Khela, H., Bettegowda, A. \& Bettegowda, C. Biomarkers and Focused Ultrasound: The Future of Liquid Biopsy for Brain Tumor Patients. \textit{J. Neurooncol.} \textbf{156}, 33--48 (2022).
\bibitem{hofmann2023} Hofmann, B., Brandsaeter, I. \O{}. \& Kjelle, E. Variations in Wait Times for Imaging Services: A Register-Based Study of Self-Reported Wait Times for Specific Examinations in Norway. \textit{BMC Health Serv. Res.} \textbf{23}, 1287 (2023).
\bibitem{panfilova2021} Panfilova, A., van Sloun, R. J. G., Wijkstra, H., Sapozhnikov, O. A. \& Mischi, M. A review on B/A measurement methods with a clinical perspective. \textit{J. Acoust. Soc. Am.} \textbf{149}, 2200--2237 (2021).
\bibitem{zhang2001} Zhang, D., Gong, X. \& Chen, X. Experimental imaging of the acoustic nonlinearity parameter B/A for biological tissues via a parametric array. \textit{Ultrasound Med. Biol.} \textbf{27}, 1359--1365 (2001).
\bibitem{tang2011} Tang, M.-X., Loughran, J., Stride, E., Zhang, D. \& Eckersley, R. J. Effect of Bubble Shell Nonlinearity on Ultrasound Nonlinear Propagation through Microbubble Populations. \textit{J. Acoust. Soc. Am.} \textbf{129}, EL76--EL82 (2011).
\bibitem{cavaro2011} Cavaro, M., Payan, C., Moysan, J. \& Baqu\'e, F. Microbubble Cloud Characterization by Nonlinear Frequency Mixing. \textit{J. Acoust. Soc. Am.} \textbf{129}, EL179--EL183 (2011).
\bibitem{overvelde2010} Overvelde, M. \textit{et al.} Nonlinear Shell Behavior of Phospholipid-Coated Microbubbles. \textit{Ultrasound Med. Biol.} \textbf{36}, 2080--2092 (2010).
\bibitem{pinton2012} Pinton, G. \textit{et al.} Attenuation, scattering, and absorption of ultrasound in the skull bone. \textit{Med. Phys.} \textbf{39}, 299--307 (2012).
\bibitem{ingard1955} Ingard, U. \& Pridmore-Brown, D. C. Scattering of Sound by Sound. \textit{J. Acoust. Soc. Am.} \textbf{27}, 1002 (1955).
\bibitem{westervelt1957} Westervelt, P. J. Scattering of Sound by Sound. \textit{J. Acoust. Soc. Am.} \textbf{29}, 199--203 (1957).
\bibitem{fatemi1998} Fatemi, M. \& Greenleaf, J. F. Ultrasound-Stimulated Vibro-Acoustic Spectrography. \textit{Science} \textbf{280}, 82--85 (1998).
\bibitem{konofagou2001} Konofagou, E., Thierman, J. \& Hynynen, K. A focused ultrasound method for simultaneous diagnostic and therapeutic applications---a simulation study. \textit{Phys. Med. Biol.} \textbf{46}, 2967--2984 (2001).
\bibitem{karjalainen1999} Karjalainen, T., Thierman, J. S. \& Hynynen, K. Ultrasound Acoustic Stimulated Emission for Controlling Thermal Surgery. in \textit{1999 IEEE Ultrasonics Symposium Proceedings} vol. 2 1397--1400 (1999).
\bibitem{pompei2002} Pompei, F. J. Sound from Ultrasound: The Parametric Array as an Audible Sound Source. PhD thesis, Massachusetts Institute of Technology (2002).
\bibitem{berktay1965} Berktay, H. O. Possible exploitation of non-linear acoustics in underwater transmitting applications. \textit{J. Sound Vib.} \textbf{2}, 435--461 (1965).
\bibitem{blackstock2000} Blackstock, D. T. \textit{Fundamentals of Physical Acoustics}. (John Wiley \& Sons, 2000).
\bibitem{chen2004} Chen, W. \& Holm, S. Fractional Laplacian Time-Space Models for Linear and Nonlinear Lossy Media Exhibiting Arbitrary Frequency Power-Law Dependency. \textit{J. Acoust. Soc. Am.} \textbf{115}, 1424--1430 (2004).
\bibitem{rosnitskiy2015} Rosnitskiy, P. B., Yuldashev, P. V. \& Khokhlova, V. A. Effect of the angular aperture of medical ultrasound transducers on the parameters of nonlinear ultrasound field with shocks at the focus. \textit{Acoust. Phys.} \textbf{61}, 301--307 (2015).
\bibitem{moffett} Moffett, M. B. \& Henriquez, T. A. Hydrophone nonlinearity measurements. \textit{J. Acoust. Soc. Am.} \textbf{72}, 1--6 (1982).
\end{thebibliography}
}


\clearpage
{\small
\begin{thebibliography}{99}
\bibitem{si-song2021} Song, J., Jung, D., Kim, J. S. \& Lee, J. Experimental evaluation of pseudo-sound in a parametric array. \textit{J. Acoust. Soc. Am.} \textbf{150}, 3787--3796 (2021).
\bibitem{si-moffett} Moffett, M. B. \& Henriquez, T. A. Hydrophone nonlinearity measurements. \textit{J. Acoust. Soc. Am.} \textbf{72}, 1--6 (1982).
\bibitem{si-karpov2003} Karpov, S., Prosperetti, A. \& Ostrovsky, L. Nonlinear wave interactions in bubble layers. \textit{J. Acoust. Soc. Am.} \textbf{113}, 1304--1316 (2003).
\bibitem{si-treeby2010} Treeby, B. E. \& Cox, B. T. k-Wave: MATLAB Toolbox for the Simulation and Reconstruction of Photoacoustic Wave Fields. \textit{J. Biomed. Opt.} \textbf{15}, 021314 (2010).
\bibitem{si-westervelt1963} Westervelt, P. J. Parametric Acoustic Array. \textit{J. Acoust. Soc. Am.} \textbf{35}, 535--537 (1963).
\bibitem{si-hamilton2008} Hamilton, M. F. \& Blackstock, D. T. (eds) \textit{Nonlinear Acoustics} (Acoustical Society of America, 2008).
\bibitem{si-clement2004} Clement, G. T., White, P. J. \& Hynynen, K. Enhanced ultrasound transmission through the human skull using shear mode conversion. \textit{J. Acoust. Soc. Am.} \textbf{115}, 1356--1364 (2004).
\end{thebibliography}
}
\end{document}